\documentclass{ws-procs975x65}

\begin{document}

\title{Conformal Window\\ and \\ Correlation Functions in Lattice Conformal QCD\footnote{Based on the work in Collaboration with K.-I. Ishikawa, Yu Nakayama and T. Yoshie ; \\ \, Talk on December 4, 2012 at SCGT 12 in Nagoya}}

\author{Y. Iwasaki}

\address{University of Tsukuba,\\
Tsukuba, 305-8577, Japan\\
$^*$E-mail: iwasaki@ccs.tsukuba.ac.jp\\
www.ccs.tsukuba.ac.jp/people/iwasaki}

\begin{abstract}
We discuss various aspects of Conformal Field Theories on the Lattice.
We mainly investigate the $SU(3)$ gauge theory with $N_f$ degenerate fermions in the fundamental representation,
employing the one-plaquette gauge action and the Wilson fermion action.

First  we make a brief review of our previous works on the phase structure of lattice gauge theories in terms of the gauge coupling constant and the quark mass. We thereby clarify the reason why we conjecture that the conformal window is $7 \le N_f \le 16$.

Secondly, we introduce a new concept,''conformal theories with IR cutof'' and
point out that any numerical simulation on a lattice is bounded by an IR cutoff $\Lambda_{IR}$.
Then we make predictions that when $N_f$ is within the conformal window, 
the propagator of a meson $G(t)$ behaves at large $t$, as $G(t) = c\, \exp{(-m_H t)}/t^\alpha$, that is, a modified Yukawa-type decay form, instead of the usual exponential decay form $\exp{(-m_H t)}$,
in the small quark mass region.
This holds on an any lattice for any coupling constant $g$, as far as $g$ is between $0$ and $g^*$, where $g^*$ is the IR fixed point.
We verify that numerical results really satisfy the predictions for the $N_f=7$ case and the $N_f=16$ case.

Thirdly, we discuss small number of flavors ($N_f=2 \sim 6$) QCD  at finite temperatures. 
We point out theoretically and verify numerically that 
the correlation functions at $T/T_c > 1$ exhibit the characteristics of the conformal function with IR cutoff, an exponential decay with power correction.

Investigating our numerical data by a new method  which we call the ''local-analysis'' of propagators,
we observe that 
the $N_f=7$ case and the $N_f=2$ at $T\sim 2\, T_c$ case are similar to each other, while the $N_f=16$  case
and the $N_f=2$ at $T= 10^2 \sim 10^5 Tc$ cases are similar to each other.

Further, we observe our data 
are consistent with the picture that the $N_f=7$ case and the $N_f=2$ at $T \sim 2\, T_c$ case are close to the meson unparticle model.
On the other hand, the $N_f=16$ case
and the $N_f=2$ at $T= 10^2 \sim 10^5 \,Tc$ cases are 
close to a free state in the $Z(3)$ twisted vacuum.
All results are consistent with naive physical intuition and give clues for long standing issues at high temperatures such as why the free energy at high temperatures does not reach the Stefan-Boltzmann ideal gas limit even at $T=100\, T_c$.

\end{abstract}

\keywords{Conformal Theories, Conformal Window, IR cutoff,  Unparticle}

\bodymatter

\section{Introduction and Summary}
Recently there has been a lot of interest in finding a conformal or near-conformal theory, which could play a key role in constructing a theory Beyond the Standard Model (BSM) of particle physics\cite{review}. 
Non-perturrbative investigation of non-supersymmetric conformal theories in four dimensional space-time 
is  also ardently desired.

We already described in the abstract partly the objectives, the target, the method and the results of this work. 
In order to avoid to repeat the same things, we will skip what was described in the abstract and will describe
remaining important matters.

The strategy we take is as follows:
We define continuum conformal theories as the continuum limit of lattice theories.
This is a constructive field theory approach.
First, we propose the properties of continuum conformal theories from knowledge of the continuum super--symmetric conformal theories and RG argument\cite{RG}.
Then, we conjecture the properties of lattice theories.

Lattice theories are defined on a Euclidian four-dimensional
lattice with lattice size $N_x, N_y, N_z, N_t$ and lattice spacing $a$.
For simplicity, we assume  $N_x=N_y=N_z=N$ and $N_t=r N$, with $r$ a constant.
The continuum limit of the theory is defined by $a \rightarrow 0$ with $N \rightarrow \infty$ 
keeping $L =N \, a$  constant. 
When  $L=$ finite, the continuum limit defines a theory with an IR cutoff.

We propose that the propagator of a meson $G(t)$ behaves at large $t$, as $G(t) = c\, \exp{(-m_H t)}/t^\alpha$ for a small enough quark mass.
We note that this form of the propagator in the coordinate representation corresponds to a cut instead of a pole in the momentum representation.

After we have verified that numerical results satisfy the predictions for the $N_f=7$ case and the $N_f=16$ case,
we propose a new method of analysis of propagators which we call the ''local-analysis of propagator'':
Writing the propagator of a meson as $G(t)=c\, \exp{(- m(t) t)}/t^{\alpha(t)}$, 
investigate the $t$ dependence of $m(t)$ and $\alpha(t)$.
The $m(t)$ and $\alpha(t)$ are reflected by the evolution of RG transformation flow and therefore contain useful information of the dynamics.

Carefully investigating the  $m(t)$ and $\alpha(t)$,
we observe that our data are consistent with the picture that the $N_f=7$ theory is close to the meson unparticle model, while the $N_f=16$ theory is close to free fermion state.
We also observe the similarity between the large  $N_f$ case and the small $N_f$ at $T \ge T_c$ case.

\section{Stage and Tools}

We employ 
the Wilson quark action and the standard one-plaquette gauge
action. 
For fermion field an anti-periodic boundary condition is imposed in the
time direction, otherwise periodic boundary conditions.
The theory is defined by two parameters; the bare coupling constant $g_0$ and the bare quark mass $m_0$.

We usually use, instead of the bare gauge coupling constant $g_0$ and the bare quark mass $m_0$,
$\beta$ defined by 
$\beta={6}/{g_0^2}$
and 
$K= 1/2(m_0a+4)$,	
which is called the hopping parameter.

In order to investigate properties of the theory in the
continuum limit,
one has to first clarify the phase structure of lattice 
QCD at zero temperature to
identify a UV fixed point and/or
an IR fixed point.
When $N_F \leq 16$, the point $g_0=0$ and $m_0=0$ is a UV fixed point. Therefore a theory governed by this fixed point is an asymptotically free theory. We restrict ourselves to the theory defined by this UV fixed point in this articles.

When there is an IR fixed point\cite{Banks1982}
 at finite coupling constant $g$ on the massless line which starts from the UV fixed point, the long distance behavior is determined by the IR fixed point. This defines a conformal theory.
The minimum number of flavors $N_f$ for the existence of such an IR fixed point is denoted by $N_f^{c}$. 
Usually $N_f^{c} \le N_f \le 16$ is called the conformal window.

The local meson operator is denoted by
$G_H(t)$.
When the theory is in the confining phase,
it decays exponentially at large $t$ 
\begin{equation} G_H(t) = c \, \exp(-m_H t)\end{equation}
for a free boundary,
where $m_H$ is the mass of the ground state hadron contained in the channel $H$.

The boundary conditions we take imply
\begin{equation} G(t) = c \, ( \exp(-m_H t) + \exp(-m_H (L_t - t)))\end{equation}
for $t \rightarrow L_t/2.$
From this behavior we extract the energy of the ground state $m_H$ in the channel $H.$

The quark mass is defined
through Ward-Takahashi identities 

\begin{equation}
m_q 
=  \frac{\langle 0 | \nabla_4 A_4 | {\rm PS} \rangle}
        {2\langle  0 | P | {\rm PS} \rangle}, 
\label{eq:quark-mass}
\end{equation}
where $P$ is the pseudo-scalar density and $A_4$ the fourth component of the
local axial vector current. Renormalization constants are suppressed, since this expression is enough for our purpose.

\section{Phase Diagram: a brief survey of our previous works}
In our earlier article \cite{iwa2004}, we investigated the phase structure of the SU(3) and SU(2) gauge theory with various numbers of flavors $N_f$ of Wilson fermions in the fundamental representation, and conjectured that the conformal window  is $7 \le N_F \le 16$ for the SU(3) and $3 \le N_F \le 10$ for the SU(2), based on numerical results for various numbers of flavors.

\def\figsubcap#1{\par\noindent\centering\footnotesize(#1)}
\begin{figure}[b]%
\begin{center}
  \parbox{2.1in}{\epsfig{figure=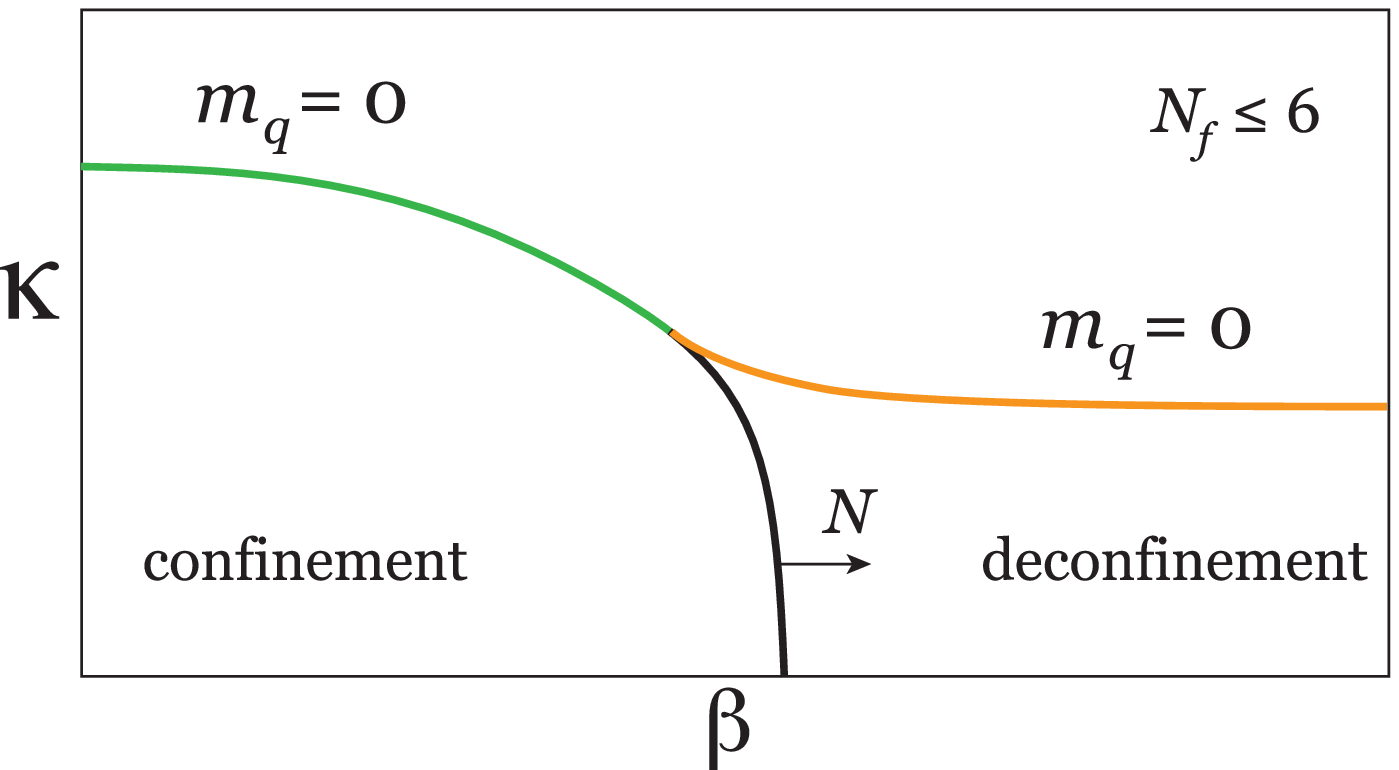,width=2in}\figsubcap{a}}
  \hspace*{4pt}
  \parbox{2.1in}{\epsfig{figure=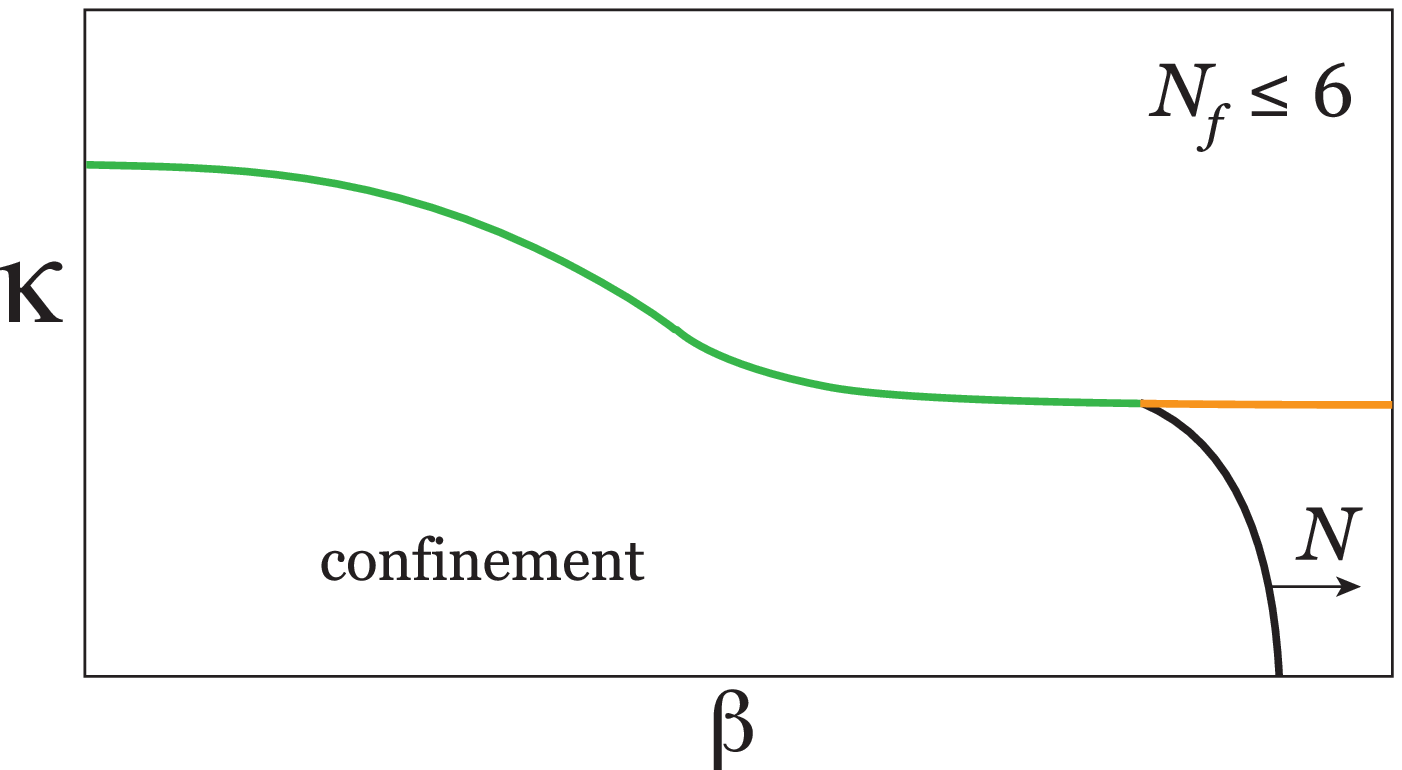,width=2in}\figsubcap{b}}
  \caption{Phase diagram for $N_f \le 6$. (a) For not so large$N$. (b) For larger $N$.}
  \label{Nf6}
\end{center}
\end{figure}
	
Although we would like to extract the phase diagram at infinite volume and at zero temperature, we have to make simulations at finite volume $N^3$ and at finite $N_t=r N$. 
Therefore the phase diagram is a three dimensional space parameterized by, $g_0$, $m_0$ and 
$N$.  Finally we have to take the limit $N$ infinity. 
  
 \subsection{$N_f \le 6$}
 
When $N$ is finite, the phase diagram is given as in Fig. \ref{Nf6}(a).
In 1996, we studied \cite{iwa1996} the finite temperature transition of QCD for $N_F$ =2, 3, 6. Our results implies that the transition in the chiral limit is of first order for $N_f= 3$ and $6$. 
The order of the chiral transition in the $N_f=2$ case is more subtle. 
We will discuss this issue later.

At $K=0$ which corresponds to the pure gauge theory, the phase transition is of first order.
Thereby the transition line connects between the critical point at $K=0$ and the chiral transition point at $K=K_c$. 

As lattice size $N$ increases, the phase transition points both at quenched QCD and at massless QCD
 move towards larger $\beta$ as shown in Fig.\ref{Nf6}(b). Eventually in the limit $N \rightarrow \infty$, all the region is covered with the confinement phase as shown in Fig.\ref{Nf6and7}(a).

 \subsection{$7  \le N_f \le 16$}
When $N$ is finite, we found that the phase diagram is given as in Fig. 2(a). 
Due to  the lack of chiral symmetry the structure of phase diagram is complicate.

\def\figsubcap#1{\par\noindent\centering\footnotesize(#1)}
\begin{figure}[b]%
\begin{center}
  \parbox{2.1in}{\epsfig{figure=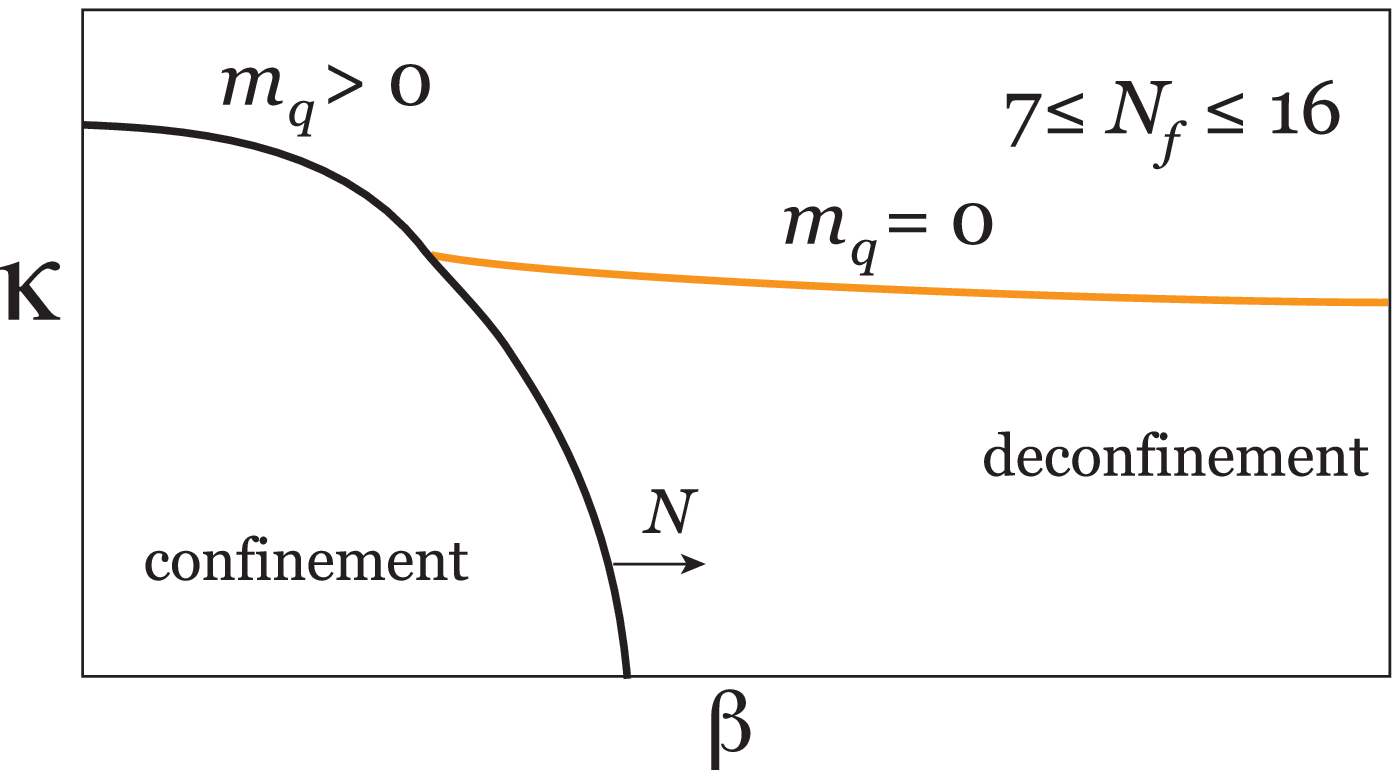,width=2in}\figsubcap{a}}
  \hspace*{4pt}
  \parbox{2.1in}{\epsfig{figure=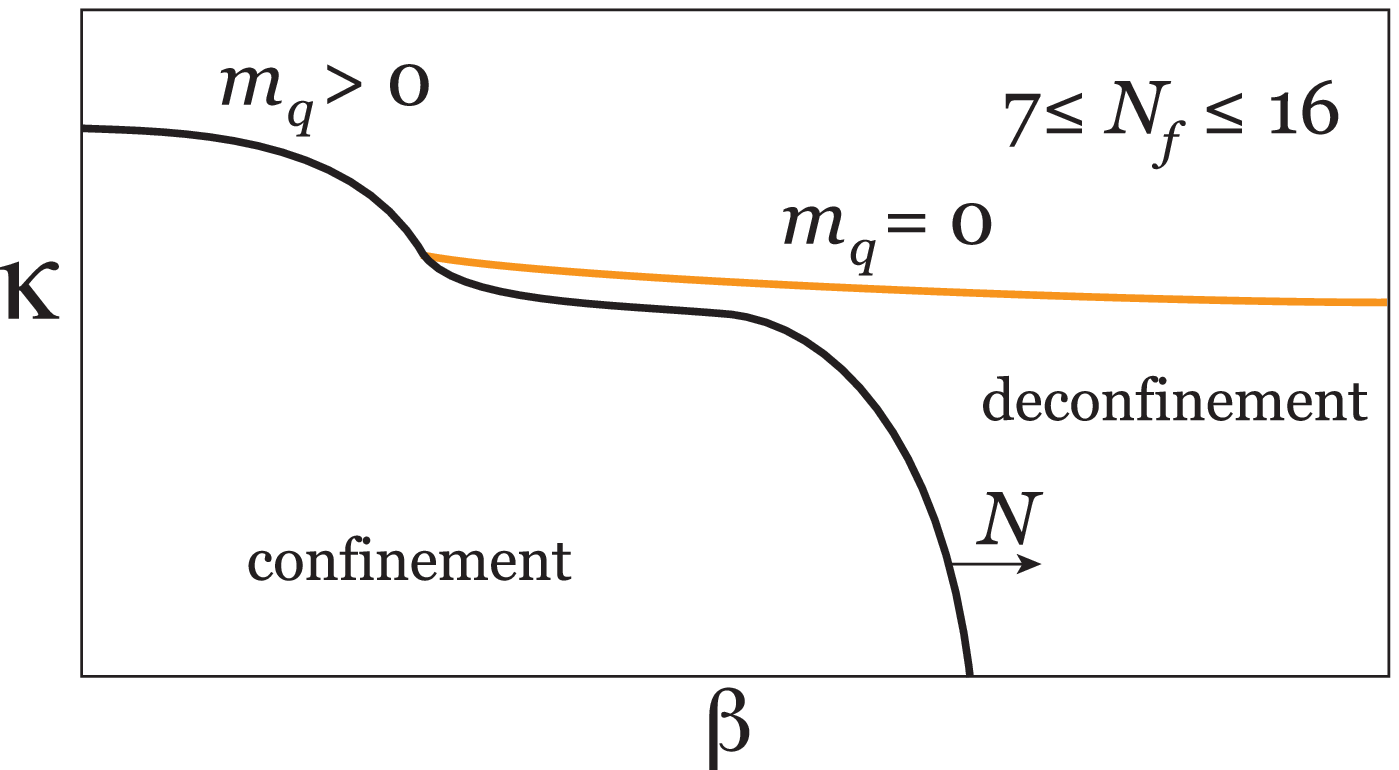,width=2in}\figsubcap{b}}
  \caption{Phase diagram for $7 \le N_f \le 16$. (a) For not so large$N$. (b) For larger $N$.}
  \label{Nf7}
\end{center}
\end{figure}

The salient facts are the following:

First, the massless line which starts the UV fixed point moves towards smaller $\beta$ and hits the bulk transition. The massless line belongs to a deconfining phase all through the region.

Secondly, in the confinement phase which exists at small  $\beta$ region there is no massless line. 
When we decrease the quark mass from the quenched QCD, we encounter
a bulk transition at finite quark mass. Thus this confinement phase is irrelevant to the continuum theory.

\def\figsubcap#1{\par\noindent\centering\footnotesize(#1)}
\begin{figure}[b]%
\begin{center}
  \parbox{2.1in}{\epsfig{figure=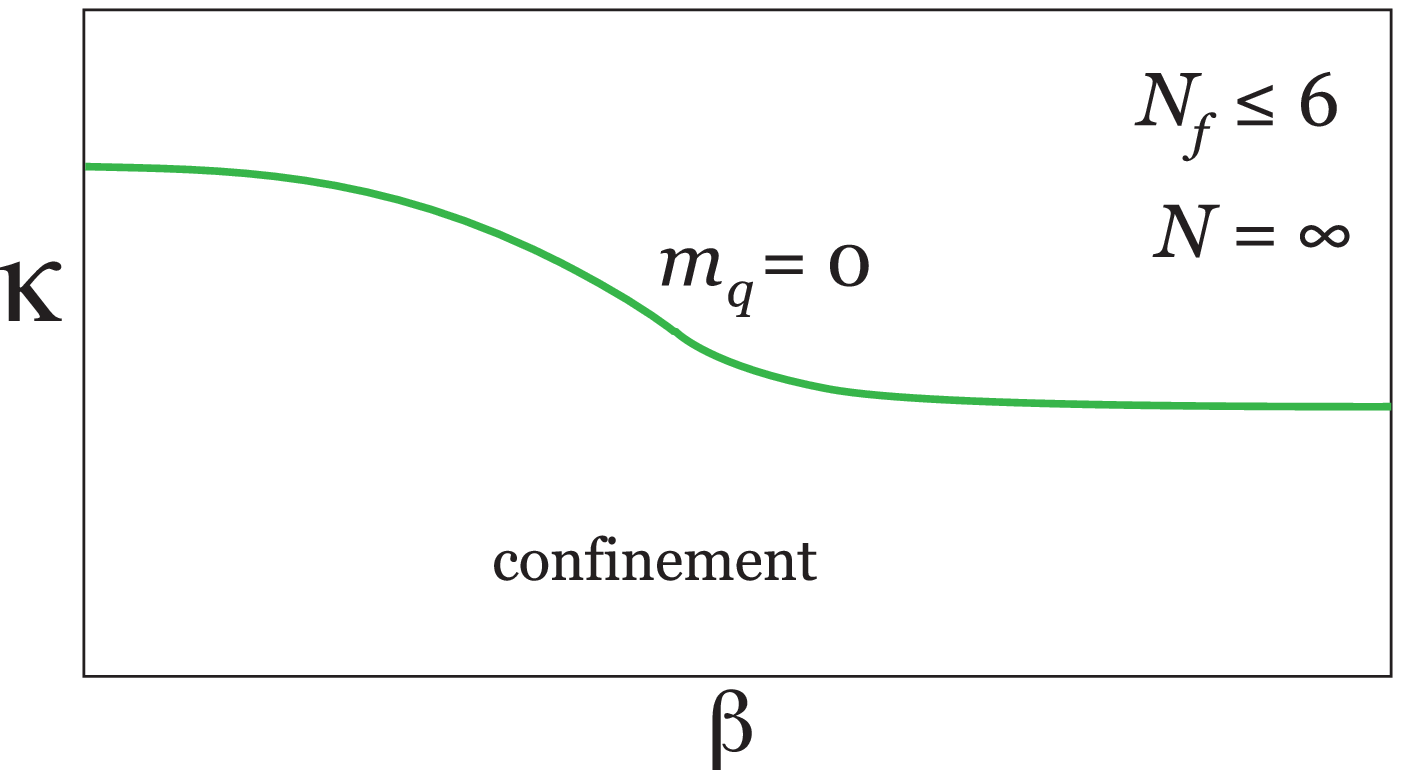,width=2in}\figsubcap{a}}
  \hspace*{4pt}
  \parbox{2.1in}{\epsfig{figure=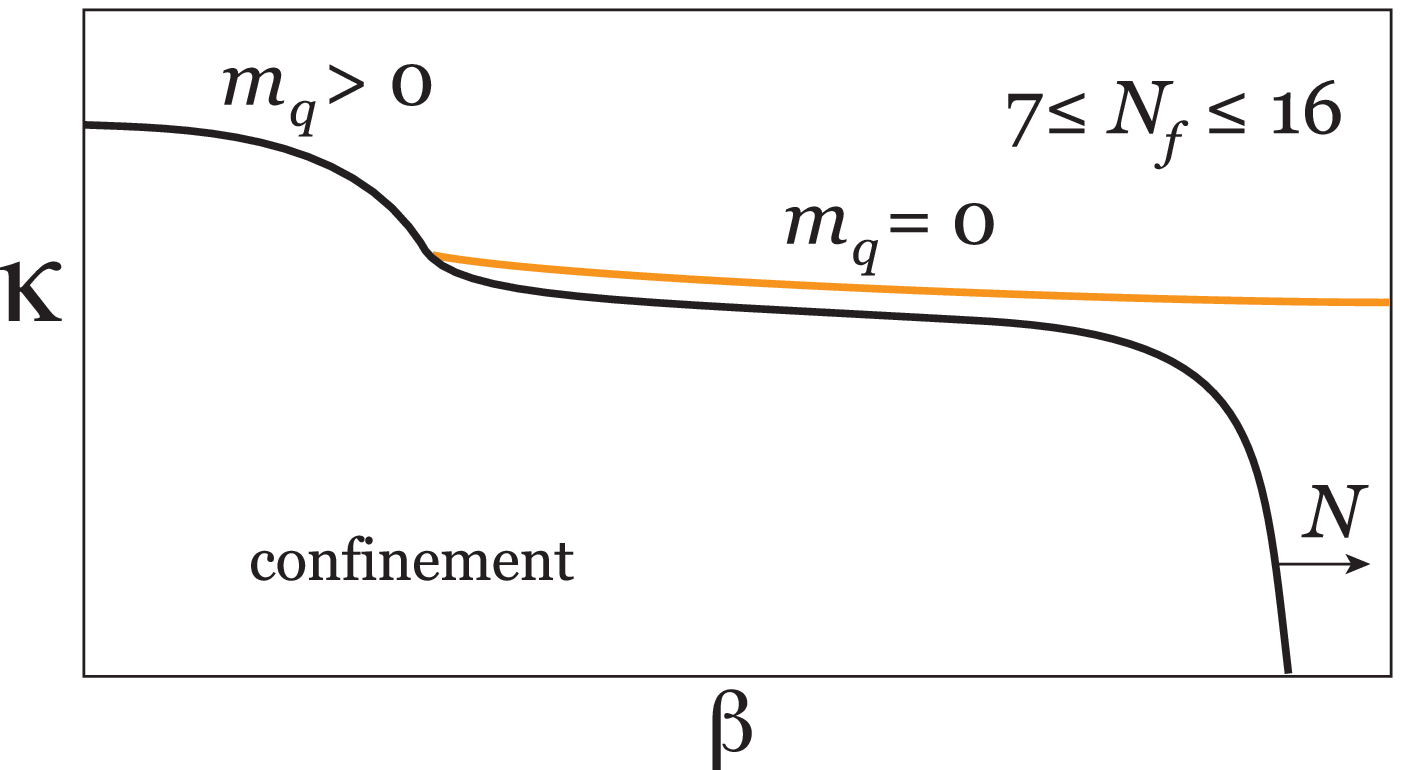,width=2in}\figsubcap{b}}
  \caption{Phase diagra for large $N$. (a) For $N_f \le 6$. (b) For $7 \le N_f \le 16.$}
  \label{Nf6and7}
\end{center}
\end{figure}

As lattice size $N$ increases the phase diagram changes as shown in Fig.\ref{Nf7}(b) and Fig.\ref{Nf6and7}(b).
It should be noted that there is still no massless line which belongs to a confinement phase.
This implies that even in the continuum limit, the massless line does not belong to the confinement phase.
On the other hand, if there would be no IR fixed point, the massless line should belong to a confining phase in  the continuum limit.

More generally, consider a set of actions of the Wilson quark action and improved gauge actions.
There is no chiral non-symmetric phase on the massless line in this set of actions. 
Assume that the quark confinement is realized for a some large lattice with some action which belongs to the set of actions. 
Then make a renormalization transformation towards the IR region.
The IR behavior does not change following the idea of RG transformation.
For the change of factor $s$, the linear size of the lattice becomes $1/s$ and the gauge coupling constant
$g$ becomes larger. Repeat this procedure.
Then finally the lattice size becomes similar to that we investigated the phase structure.
Since the IR behavior should not change, there must exist the confinement phase on a small lattice
also. However, there is no confinement region on a small lattice within the set of actions.
Thus we reach contradictory.
Therefore we may conclude the quark is not confined in this set of actions.

From this reasoning we conjecture that the conformal window is $7 \le N_f \le 16$.


\section{Strategy for Identification of the Conformal Window}
The logics in the previous work is an indirect way to identify the conformal window.
Certainly more direct ways are desirable.
The most direct and solid way is to identify the IR fixed point.
However, for small $N_f$ the IR fixed point exists deep in strong coupling region, and it becomes difficult to identify the IR fixed point.

An alternative way is to find directly the characteristics of conformal theories. This is the way we take in this work.

\section{Conformal Field Theories}
In this section we address ourselves to a basic and fundamental issue of what happens 
when the number of flavor $N_f$ is within the conformal window.

The strategy we take is as follows:
We define continuum conformal theories as the continuum limit of lattice theories.
First, we propose the structure of continuum conformal theories from knowledge of the continuum super--symmetric conformal theories and RG argument.
Then, we conjecture the structure of conformal theories on the lattice. 

\subsection{Continuum Conformal Field Theories}
We discuss separately the $L=\infty$ case and the  $L=$ finite case,
since there exists a clear difference between the two cases.

\def\figsubcap#1{\par\noindent\centering\footnotesize(#1)}
\begin{figure}[b]%
\begin{center}
  \parbox{2.1in}{\epsfig{figure=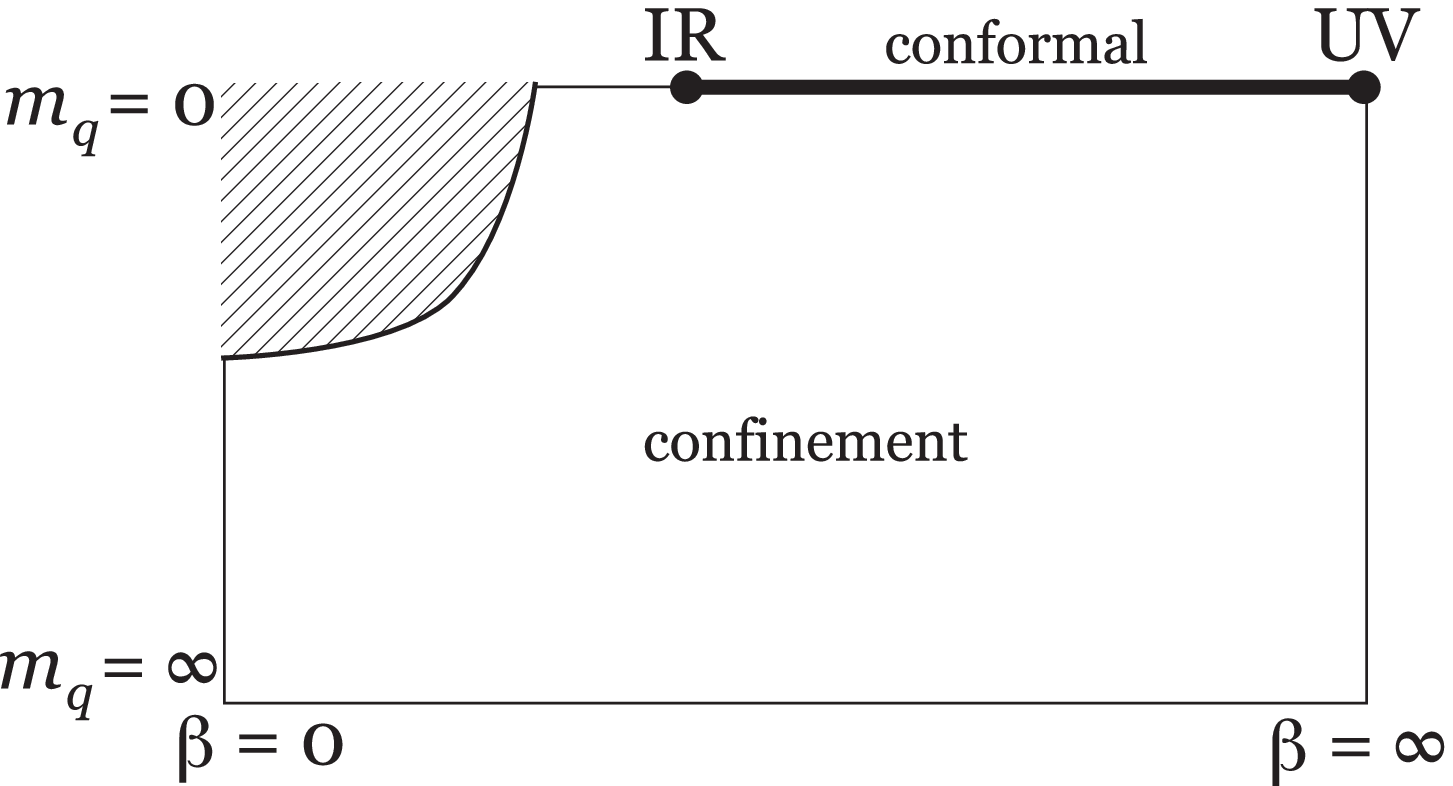,width=2in}\figsubcap{a}}
  \hspace*{4pt}
  \parbox{2.1in}{\epsfig{figure=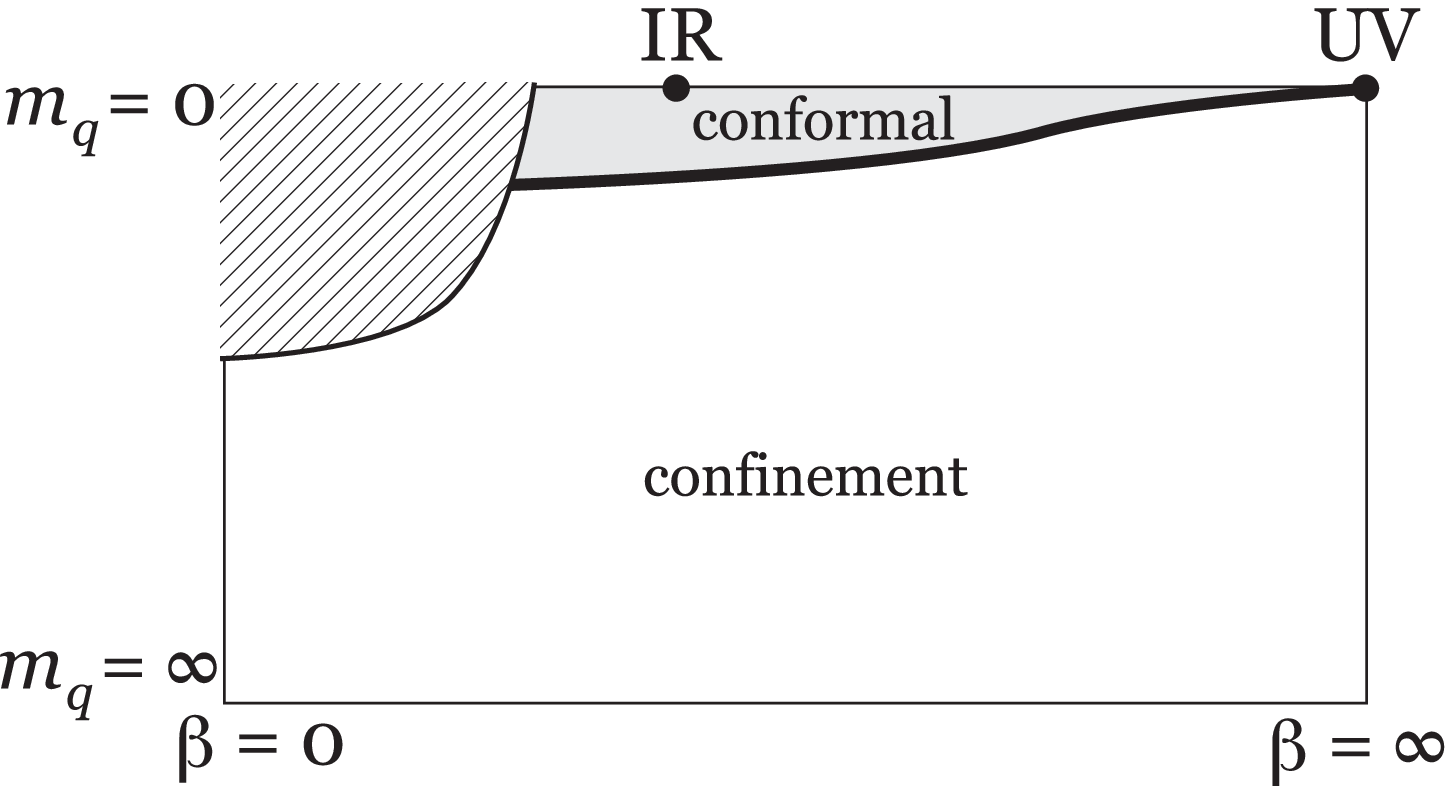,width=2in}\figsubcap{b}}
  \caption{Phase diagra for large $N=\infty$. (a) $\Lambda_{IR}=0$. (b) $\Lambda_{IR}=$ finite}\label{PDIR}
\end{center}
\end{figure}

\subsubsection{The case $L=\infty$}
This defines a continuum theory on $R^4$ Euclidean four-dimensional space, which corresponds to the flat infinite space at zero temperature.
In this case the IR cutoff $\Lambda_{IR}$ is zero. 

Therefore the relevant quantity is only the running coupling  constant $g(\mu)$ and
there are no physical quantities which have physical dimensions. 

The ''conformal region'' exists only on the massless line.
All other region is the ''confinement region''.
The phase structure is shown in Fig.\ref{PDIR}(a).
In super-symmetric theories this kind phase structure is known.
The shaded region in the strong coupling region does not exist in the $\beta - m_q$ plane.
When the phase structure is described in terms of $\beta - K$, the corresponding phase belongs to a region for Wilson doubles (Aoki phase). Therefore when it is mapped to the $\beta - m_q$ planes, the region corresponding to the shaded one does not exist.

The propagator on the massless quark line takes the form
\begin{equation} G(t) = c \, \frac {1}{t^\alpha}, \end{equation}
since, as mentioned, there is no physical quantities with physical dimensions.

If we take the coupling constant $g=g^{*}$ at UV scale $\mu$, 
\begin{equation} \alpha(t)=3 - 2 \gamma^{*},\end{equation}
with $\gamma^{*}$ being the anomalous mass dimension $\gamma$ at $g=g^{*}$.
The theory is scale invariant (and shown to be conformal
invariant within perturbation theory \cite{Polchinski:1987dy}.
See also e.g. Nakayama\cite{Nakayama:2010zz} and references therein from AdS/CFT approach).
In general, the power $\alpha$ varies with $t$:
\begin{equation}\alpha = \alpha(t).\end{equation}

When $t$ is small, the correlation function is governed by the UV fixed point. There the ''meson state'' is a free quark and an anti-quark state. Thus the power is given by
\begin{equation}
\alpha(t) =3.
\end{equation}

When $t$ is large, it is governed by the IR fixed point. 
The exponent is given by
\begin{equation} \alpha(t)=3 - 2 \gamma^{*}.\end{equation}

The exponent of the power $\alpha$ in IR is given by the renormalization equation or derived just by dimension counting.
 
\subsubsection{The case $L=$ finite}
This defines a continuum theory on $T^4$ Euclidean four-dimensional space, which corresponds to a compact $T^3$ space at finite temperature.
In this case the IR cutoff $\Lambda_{IR}=1/L$ is finite.

The renormalization transformation to IR does stop to evolve at the scale of IR cutoff $\Lambda_{IR}$.
Thus the relevant physical quantities we may choose are
\begin{equation}
\Lambda_{CFT}, \Lambda_{IR}, m_H
\end{equation}
Here $\Lambda_{CFT}$ is a scale parameter for the transition region from UV to IR, and
$m_H$ is a generic mass of a meson such as the pion and the $\rho$ meson.

Now we make a proposal.
The propagator of a meson $G(t)$ behaves at large $t,$  as
 \begin{equation}
G(t) = c\\ \frac {\exp(-mt)}{t^\alpha},
\label{yukawa type}
 \end{equation}
 that is, a modified Yukawa-type decay form in the region
\begin{equation} 
m_H   \leq c \,  \Lambda_{IR},
\label{critical mass}
\end{equation}
 instead of the usual behavior of the propagator in the ''confinement region'',
\begin{equation} G(t) = c\\ {\exp(-mt)}. \end{equation}

Here $\Lambda_{IR}$ is an IR cutoff, $m_H$ is a generic mass of a hadron such as the pion mass., and $c$ is a constant of order 1.  The conformal region is depicted in Fig.\ref{PDIR}(b).
The relation eq.(\ref{critical mass}) reduces to
\begin{equation} m_q   \leq c \,  (1/L)^{(1+\gamma^{*})},\label{critical quark mass}\end{equation}
using the relation\cite{miransky}\label{scaling}
\begin{equation}m_H = c \, m_q^{1/(1+\gamma^{*})}.\end{equation}

The phase structure with  $L=$ finite smoothly approaches the case  $L=\infty,$ as $L$ varies to $\infty.$
The form of the propagator also smoothly becomes \begin{equation} G(t) = c\\ \frac {1}{t^\alpha},\end{equation} 
in the massless limit.

As is the case in the $\Lambda_{IR}=0$ case, the power $\alpha$ varies with $t$:
\begin{equation}\alpha = \alpha(t).\end{equation}

When $t$ is small, 
\begin{equation}  \alpha(t) =3. \end{equation}

On the other hand, when $t$ is large $(t \, \gg \,  1/\Lambda_{CFT}),$
the behavior of $\alpha(t)$ is complicate:
When $m$ is small enough and if $m\, t \ll 1,$
the power is given by
\begin{equation} \alpha(t)=3 - 2 \gamma^{*}.
\label{anomalous mass}
\end{equation}
If we take larger $t$ such that $m\, t \gg 1$, the $\alpha(t)$ takes a value which depends on the dynamics of the theory.
We give explicit examples when we discuss some models.

The concept of ''conformal theory with IR cutoff'' and ''conformal region'' determined by equations~~(\ref{yukawa type}) and (\ref{critical mass}) are our key proposals.

\subsection{Conformal Field Theories on the Lattice}
We are able to perform numerical simulations only on a finite size lattice.
Thus the system is automatically bounded by the IR cutoff $\Lambda_{IR} \sim 1/(N a).$
In this case,
the relevant physical parameters we may use  are $\Lambda_{CFT}$,  $\Lambda_{IR}$ and $m_H.$

Our main target is first to verify the transition of a meson propagator from an exponential decay form
to a modified Yukawa-type, that is, an exponential form with power correction given by eq.~(\ref{yukawa type}), at the critical hadron mass given by eq.~(\ref{critical mass}).

When  the lattice size $N_1$ increased to $N_2=s \, N_1$  with $\beta$ being fixed,
then 
 \begin{equation}m_H^{critical}(\beta, N_2)= 1/s \, m_H^{critical}(\beta, N_1).\end{equation}
The critical hadron mass decreases.
Thus we have to carefully choose the simulation parameters in order to find the conformal region and the Yukawa-type propagators.

\section{Numerical Simulations}
\subsection{Algorithm and Parameters}

\begin{table}
\tbl{Parameters of Simulations }
{\begin{tabular}{@{}cccccccccccc@{}}
\toprule
Nf &$\beta$ & K & mq &Nf & $\beta$ & K & mq&Nf&$\beta$&K&mq\\
7 &6.0 &  & &16 & 11.5& & &  2& 6.5 & & \\
&& 0.1400\hphantom{00} & \hphantom{0}0.22 & & &  \hphantom{0}0.126 & 0.22& & & 0.146 & 0.035\\
&& 0.1446\hphantom{00} & \hphantom{0}0.084 & & &  \hphantom{0}0.127 & 0.19& & 10.0& &\\ 
&& 0.1452\hphantom{00} & \hphantom{0}0.062 & &  &\hphantom{0}0.130 & 0.10& && 0.135 & 0.028\\
&& 0.1459\hphantom{00} & \hphantom{0}0.045 &  & & \hphantom{0}0.1315 & 0.055& & 15.0& &\\
&& 0.1472\hphantom{00} & \hphantom{0}0.006 &  &  & \hphantom{0}0.13322 & 0.003& & &0.130 & 0.046\\
\end{tabular}}
\label{tbl1}
\end{table}

We have made simulations mainly for the cases $N_f=7$ and $N_f=16$, which are, we think, the boundaries of the conformal window. 
The algorithm we employ is the blocked HMC algorithm \cite{Hayakawa:2010gm} for $N_f=2\, N$ and the RHMC algorithm \cite{Clark:2006fx} for $N_f=1,$ in the case $N_f=2\, N +1$.
The parameters we have simulated are listed in Table1. (The parameters for $N_f=2$ at high temperatures are also listed.)
We have chosen the hopping parameter $K$ in such a way that the quark mass takes the value from 0.25 to 0.0. It might be worthwhile to mention that we are able to take $K$ which corresponds to a negative quark mass crossing the zero quark mass, since the phase is a deconfining phase and therefore the pion mass is not zero there. The coupling constant of the IR fixed point for $N_f=16$ is $\beta=11.48$ in one-loop approximation. This is the reason why we choose $\beta =11.5$ for $N_f=16.$

The lattice sizes we choose are $8^3\times 32, 16^3\times 64$ and $24\times 96.$
The run-parameters are chosen such that the acceptance of the global metropolis test is about $70\%.$
The statistics is in general 1,000 MD trajectories for thermalization and 1,000 MD trajectories for measurement.

\section{Numerical Results}
Let  us first discuss the results for $N_f=7$ case on a $16^3\times 64$ lattice. 
Fig.\ref{exp vs yukawa} shows 
the effective mass plots
for pions with the local-sink local-source (black squares), 
local-sink doubly-smeared-source (red circles) and 
local-sink doubly-wall-source (green triangles) cases.
The effective masses in the relatively large quark mass case shown on the left panel, $m_q=0.25$($K=0.1400$),
clearly show the plateau at $t=22\sim 31$. On the other hand, in the $m_q=0.045$ ($K=0.1459$) case shown on the right panel,
there is no plateau up to $t=31$. The effective mass is slowly decreasing.

\def\figsubcap#1{\par\noindent\centering\footnotesize(#1)}
\begin{figure}[b]%
\begin{center}
  \parbox{2.1in}{\epsfig{figure=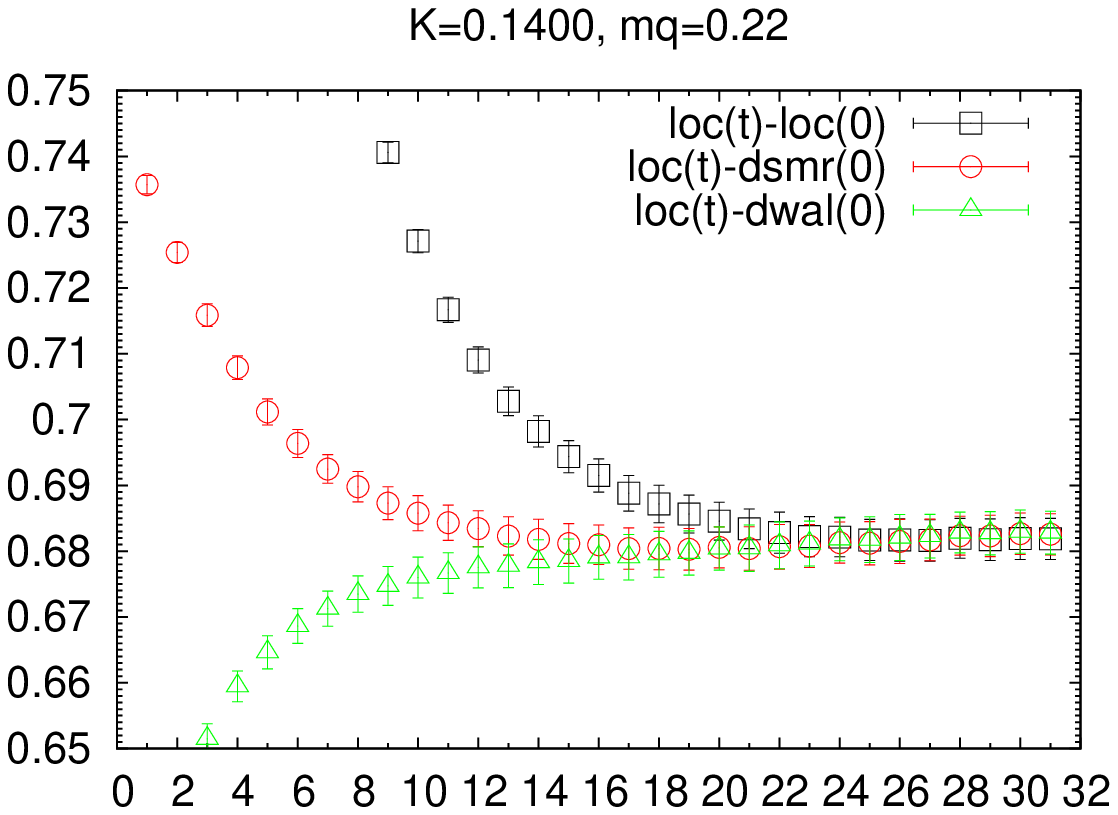,width=2.5in}\figsubcap{a}}
  \hspace*{10pt}
  \parbox{2.1in}{\epsfig{figure=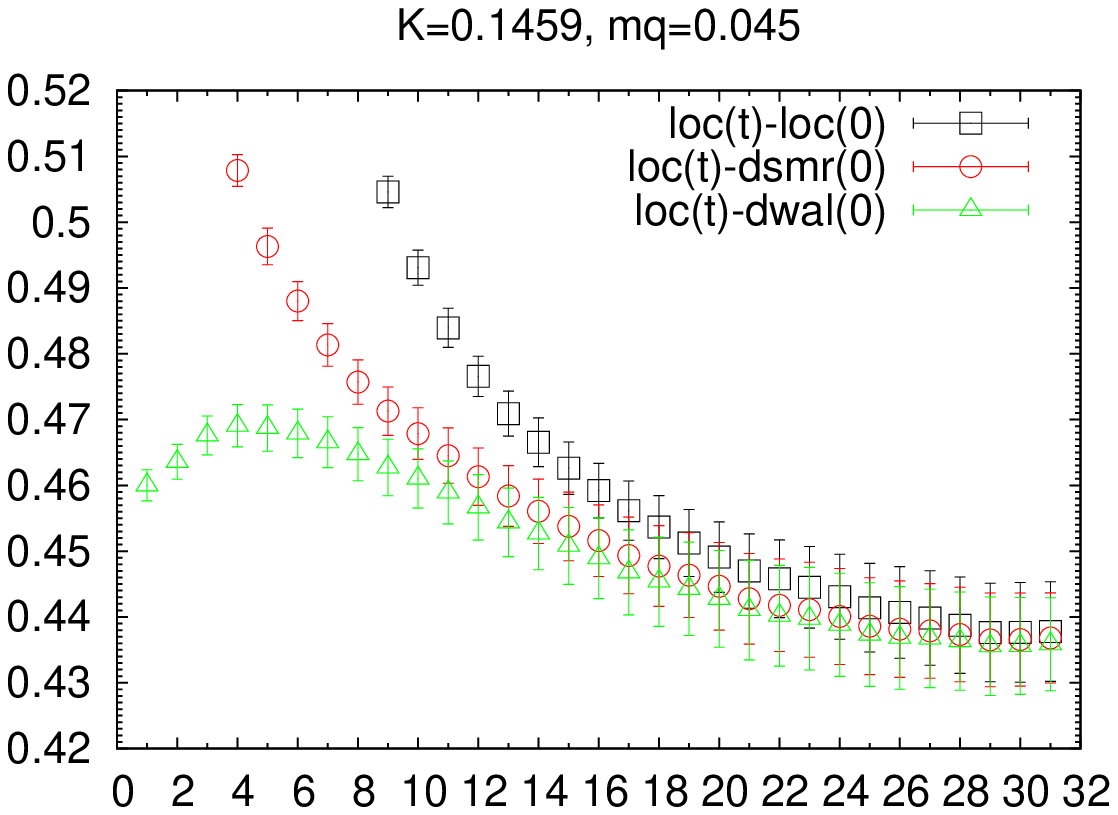,width=2.5in}\figsubcap{b}}
  \caption{The asymptotic behavior of propagators in the $N_f=7$ case on a $16^3\times 64$ lattice at $\beta=6.0$.  (a) Exponential decay form.(b) Modified Yukawa-tye decay form.}%
  \label{exp vs yukawa}
\end{center}
\end{figure}

\def\figsubcap#1{\par\noindent\centering\footnotesize(#1)}
\begin{figure}[b]%
\begin{center}
  \parbox{2.1in}{\epsfig{figure=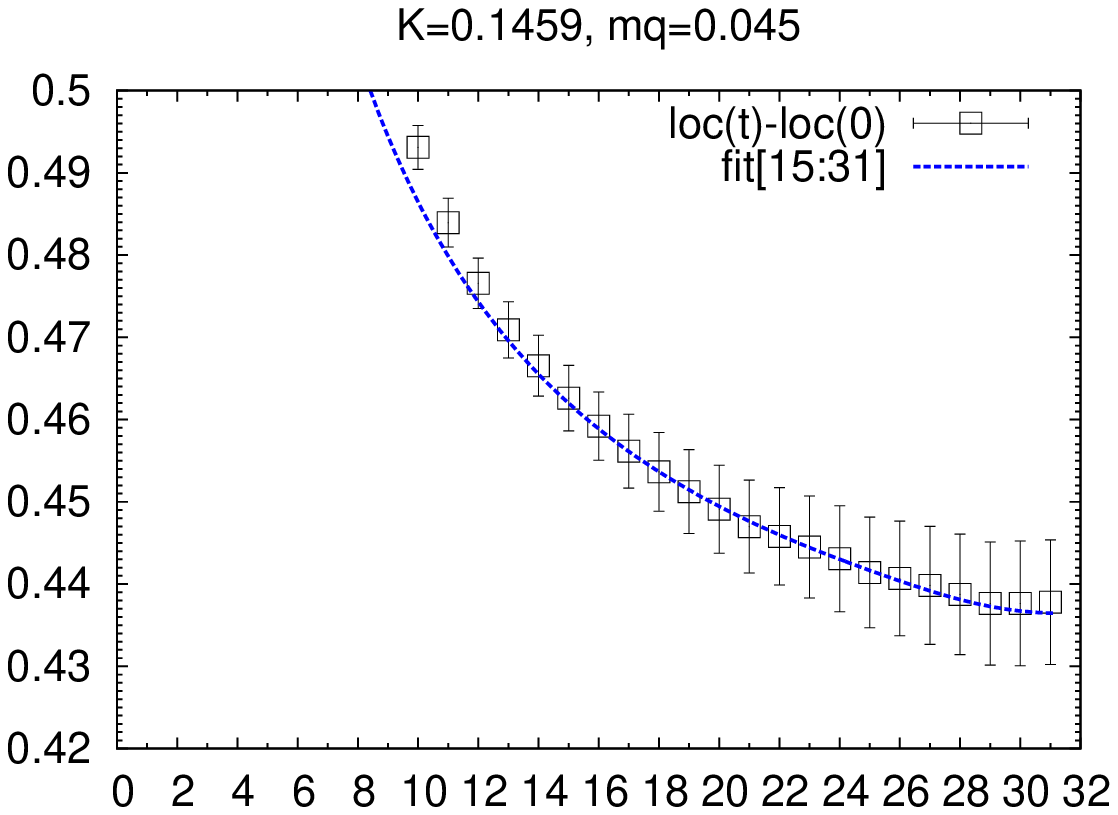,width=2.5in}\figsubcap{a}}
  \hspace*{4pt}
  \parbox{2.1in}{\epsfig{figure=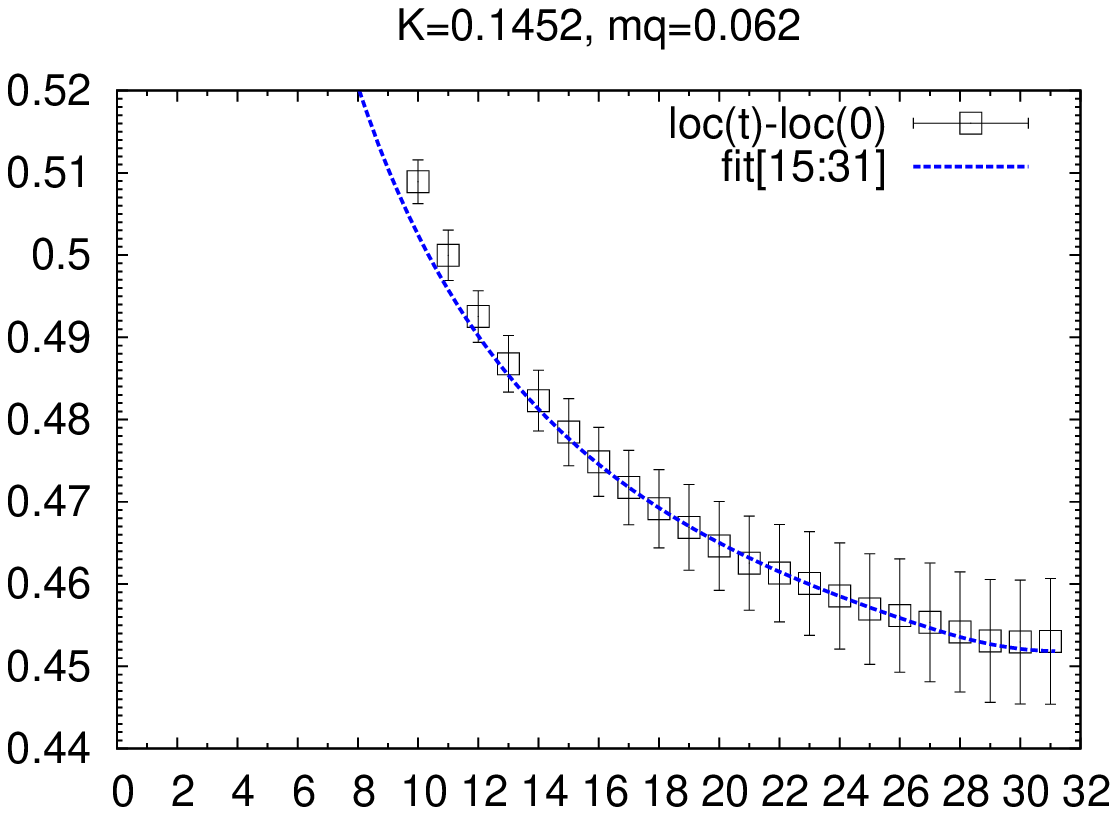,width=2.5in}\figsubcap{b}}
  \caption{Two examples of modified Yukawa-type decay form in the $N_f=7$ case on a $16^3\times 64$ lattice at $\beta=6.0$ (a) K=0.1459. (b) K=0.1452.}%
  \label{expnf7}
\end{center}
\end{figure}

\def\figsubcap#1{\par\noindent\centering\footnotesize(#1)}
\begin{figure}[b]%
\begin{center}
  \parbox{2.1in}{\epsfig{figure=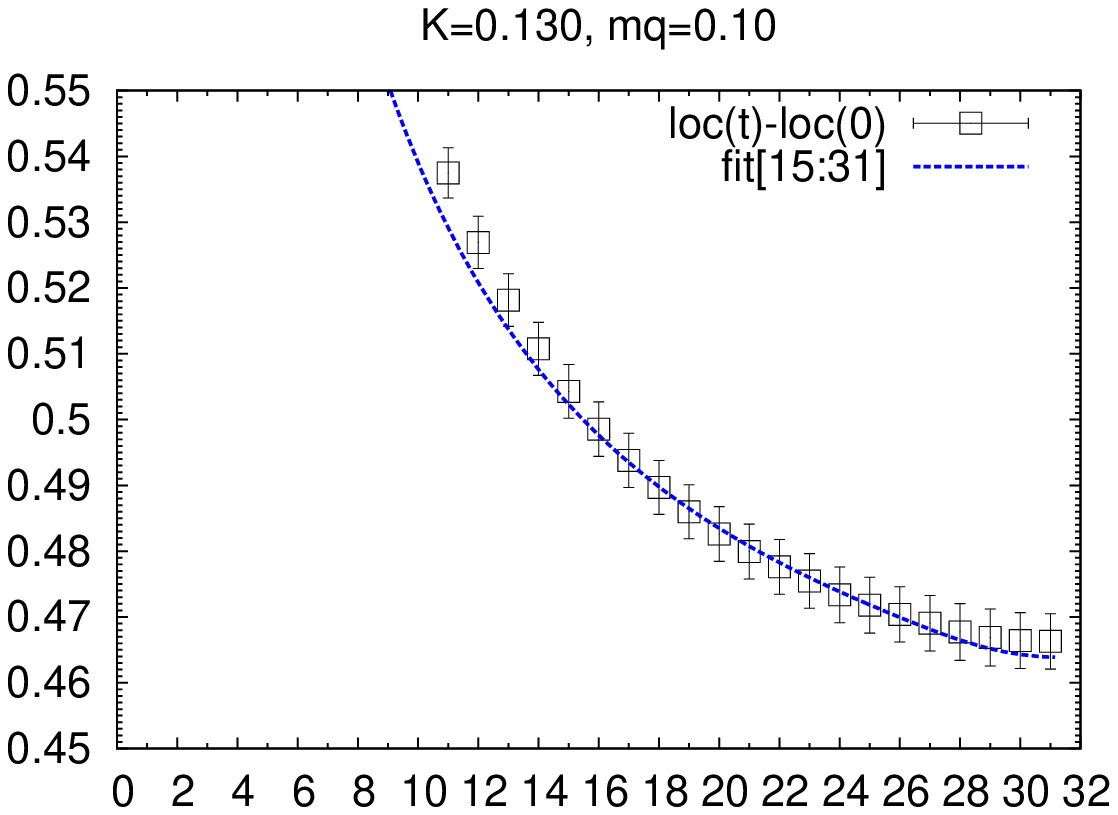,width=2.5in}\figsubcap{a}}
  \hspace*{4pt}
  \parbox{2.1in}{\epsfig{figure=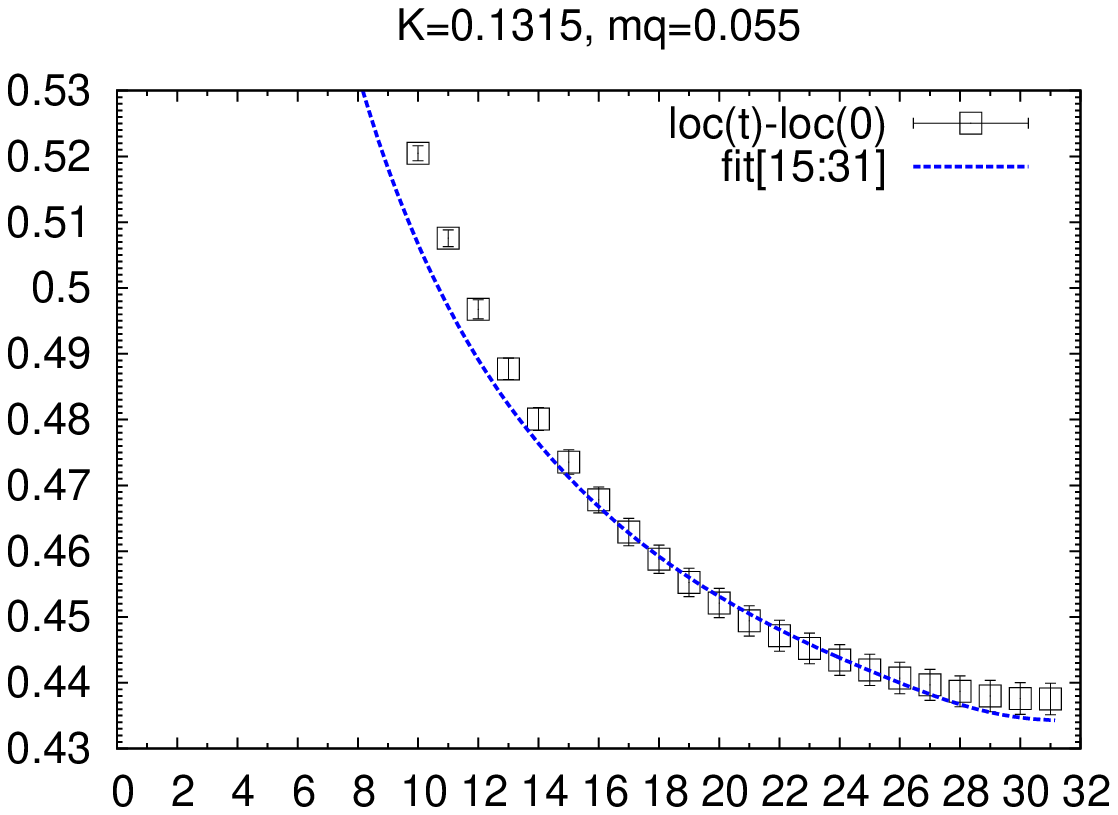,width=2.5in}\figsubcap{b}}
  \caption{Two examples of modified Yukawa-type decay form in the $N_f=16$ case on a $16^3\times 64$ lattice at $\beta=11.5.$ (a) K=0.130. (b) K=0.1315.}
    \label{exp-nf16}
\end{center}
\end{figure}

The effective masses for all cases with $m_q \le 0.084$ ($K \ge 0.1446$) exhibit similar behavior to $K=0.1459$.
We have conjectured that propagators decay exponentially with power correction.
Thus, to verify the conjecture,
we fit the local-local data with the Yukawa-type for the fitting range $t=[15:31].$
The fits  for $K=0.1459$ and $K=0.1452$ are depicted in Fig.\ref{expnf7}.
The both fits well reproduce the data with range $t=[15:31]$ with $\alpha=0.8(1).$

In the case of $N_f=16,$ also, the effective masses on a $16^3\times 64$ lattice 
exhibit the modified Yukawa-type for all cases when $K \ge 0.130$.
The results of the fit to the local-local data for $K=0.130$ and $K=0.1315$
with the Yukawa-type for the fitting range $t=[15:31]$
are plotted in Fig.\ref{exp-nf16}.
The fits well reproduce the data. However, if we carefully compare the fits of $N_f=7$ and $N_f=16$ cases, the fit of $N_f=16$ case at large $t$ is not so good as $N_f=7$ case. We will discuss this point later.

The data on a $24^3\times 96$ lattice for both cases of $N_f=7$ with $K \ge 0.1459$ and $N_f=16$ with
$K \ge 0.130$ exhibit the modified Yukawa type decay.

Thus we have verified that the propagators decay with the Yukawa-type decay form  for the $N_f=7$ and $N_f=16$ cases, for the small quark mass.
In term of the pion mass the critical mass can be expressed as $m_H\le c\, \Lambda_{IR}$.
Here $m_H$ is the pion mass and we take $\Lambda_{IR}=2\pi\,(N^3 \times N_t)^{1/4}.$
There is no {\it a priori} definition of $\Lambda_{IR}$ in the case of asymmetric lattices. This is one of choices.
With this definition,
we estimate the constant $c$ is $c\sim 2.16$ for $N_f=7$, and $c\sim 1.94$ for $N_f=16.$

\section{Local-analysis of Propagator}
As we have verified our conjecture that the propagators decay with the Yukawa-type decay form
for the small quark mass,
the next issue we would like to solve is 
what kind of theory is defined.
To do so,
we propose a new method of analysis of propagators which we call the ''local-analysis'' of propagators:
Writing the propagator of a meson as 
\begin{equation}
G(t)=c\, \exp{(- m(t) t)}/t^{\alpha(t)}
\end{equation}
investigate the $t$ dependence of $m(t)$ and $\alpha(t)$.
The $m(t)$ and $\alpha(t)$ are reflected by the evolution of RG transformation flow and therefore contain useful information of the dynamics.

We depict  $m(t)$ and $\alpha(t)$ in Figs.9$ \sim$ 19 for various cases.

First, at small $t$ (we disregard the $\alpha(t)$ and $m(t)$ at $t=1$ and $2$, since they are affected
by the boundary effect),
\begin{equation} 
\alpha(t) \sim 3.0
\end{equation} 
 and
 \begin{equation}
m(t) \sim 2\, m_q
\end{equation}

are well satisfied, when $m_q \ge 1/N$.

As $t$ increases, $m(t)$ increases from $2 \, m_q$ to $m_H$.
The behavior of $\alpha(t)$ is more complex and it reflects the dynamics.

In the $N_f=7$ case on the $16^3\times 64$ lattice, the $\alpha(t)$ at large $t$ is relatively stable and it seems that $\alpha(t)$ shows a plateau at $t=15 \sim 31$, although it is noisy.
The statics of the data on the $24\times 96$ lattice is not enough to see the $\alpha(t)$, as it requires 
more statistics than the Yukawa-type fit with a wider range of $t.$ We are storing configurations.

The $t$ dependence of  $\alpha(t)$ in the $N_f=16$ case is quite different from that of
the $N_f=7$ case.
In the $N_f=16$ case, although there is no plateau structure in $\alpha(t)$,
there is a shoulder with a value around $1.5$ at $t= 12\sim 24$.on the $16^3\times 64$ lattice.
On the $24\times 96$ lattice (the statistics is enough, since the $N_f=16$ case is much faster than
the $N_f=7$ case), the $t$ dependence of  $\alpha(t)$ is different each other. In the case $K=0.130,$ 
the shoulder becomes very wide down to $t \sim 32.$ In the case $K=0.1315$ the shoulder is clearly 
seen for $t= 12 \sim 25$. on the other hand, the shoulder disappears and the plateau at $\alpha \sim 0.5$
is seen for $t \ge 24$ in the case $K=0.13322.$

In order to find what kind of theory is defined, 
we will investigate some models which we consider are good candidates to compare with the theory.
To do so,
we find  it is better to discuss conformal theories at high temperatures before that.
Therefore we examine conformal theories at high temperatures in the next section.

\section{Conformal Field Theories at Finite Temperature}
Now let us consider lattice QCD at high temperatures
for small $N_f$ ($2 \le N_f \le 6$) where the chiral phase transition occurs at $T_c.$
Note that the system is bounded by the infrared cutoff $\Lambda_{IR} =1/T.$
The continuum limit of the theory at finite temperature is defined by $a \rightarrow 0$ with $N \rightarrow \infty$ and 
$N_t \rightarrow \infty$ keeping $N a = L$ and $N_t a = 1/T$ constant. 
Taking the limit $L \rightarrow \infty$,
the gauge theory at finite temperature $T$ is defined.

We first consider the case that the renormalized quark mass is zero.
Then the renormalized coupling constant is the only relevant variable in the theory. 
%
A running coupling constant $g(\mu; T)$  at temperature $T$ can be defined as in the case of $T=0$.
The following discussion can be applied to any method of the definition of the running coupling constant $g(\mu; T)$. 
One way is the Wilson RG method.

Although in several literatures (see e.g. ref. \refcite{karsch})
the running coupling constant $g(\mu; T)$ was discussed, it was mainly the UV behavior.
Up to now it seems that it has not been discussed systematically the IR behavior.


In the UV region  the running coupling constant at finite $T$ is essentially as the same as that of $T=0.$
Since the theory is asymptotically free, 
at small $g$ region the following relation 
\begin{equation}g^2(\mu;T)= g^2 (\mu; T=0) +c g^4 (\mu; T=0),\end{equation}
with $c$ constant, is satisfied. The constant $c$ depends on both of the schemes of $g (\mu; T=0)$ and $g(\mu; T)$ in addition to $N_f.$
It might be worthwhile to stress that the leading term is universal in the limit
$g \rightarrow 0$.

However, in the IR region $g(\mu; T)$ may be quite different from $g(\mu;T=0)$, since the IR cut-off $\Lambda_{IR}$ in the time direction is $1/T$. This is quite different from the $T=0$ case.

Furthermore, the fact that
when $T/T_c  > 1$, the quark is not confined, implies that the running coupling constant $g(\mu; T)$ cannot be arbitrarily large.This means that there is a fixed point in the $\beta$  function when $T/T_c  > 1$.


Let us increase the temperature gradually from $T=0$.

\def\figsubcap#1{\par\noindent\centering\footnotesize(#1)}
\begin{figure}[b]%
\begin{center}
  \parbox{2.1in}{\epsfig{figure=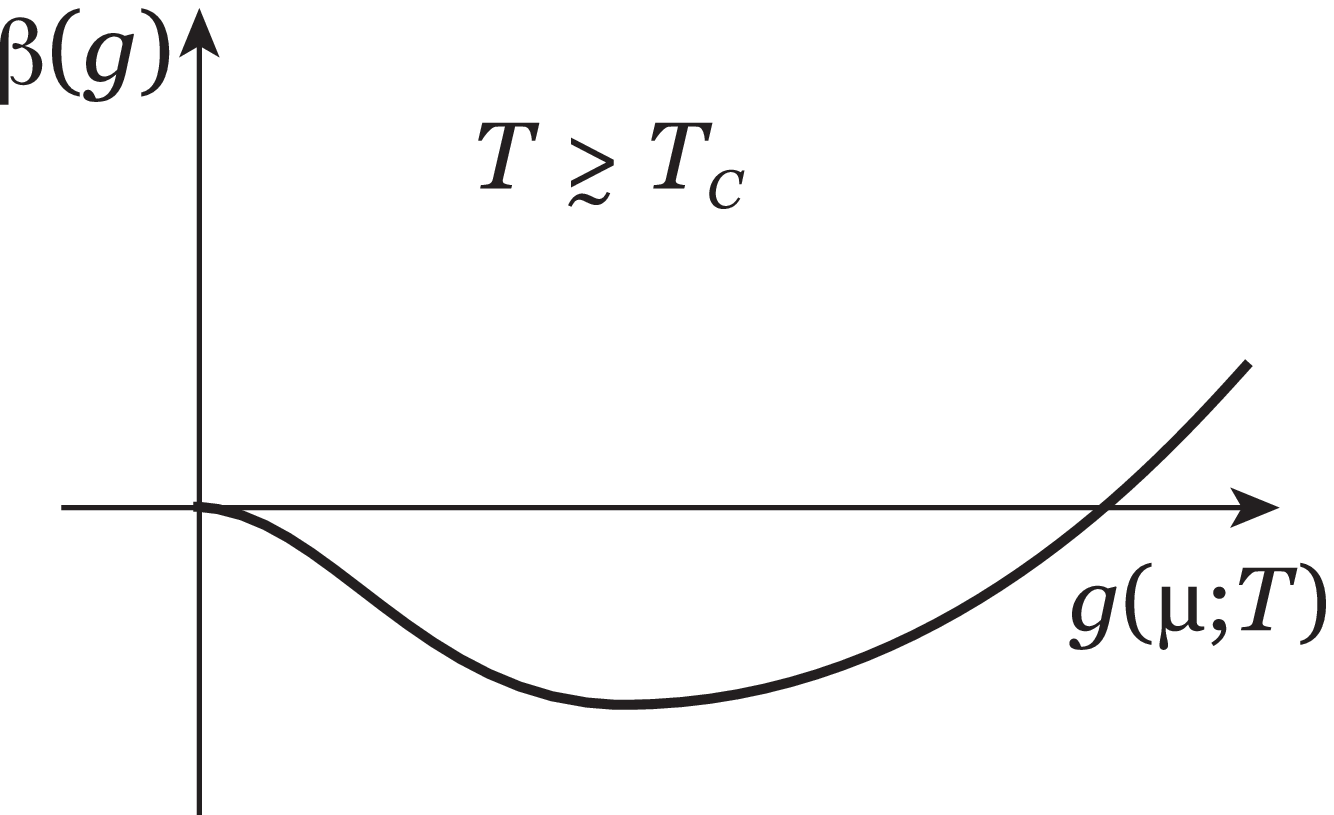,width=2in}\figsubcap{a}}
  \hspace*{4pt}
  \parbox{2.1in}{\epsfig{figure=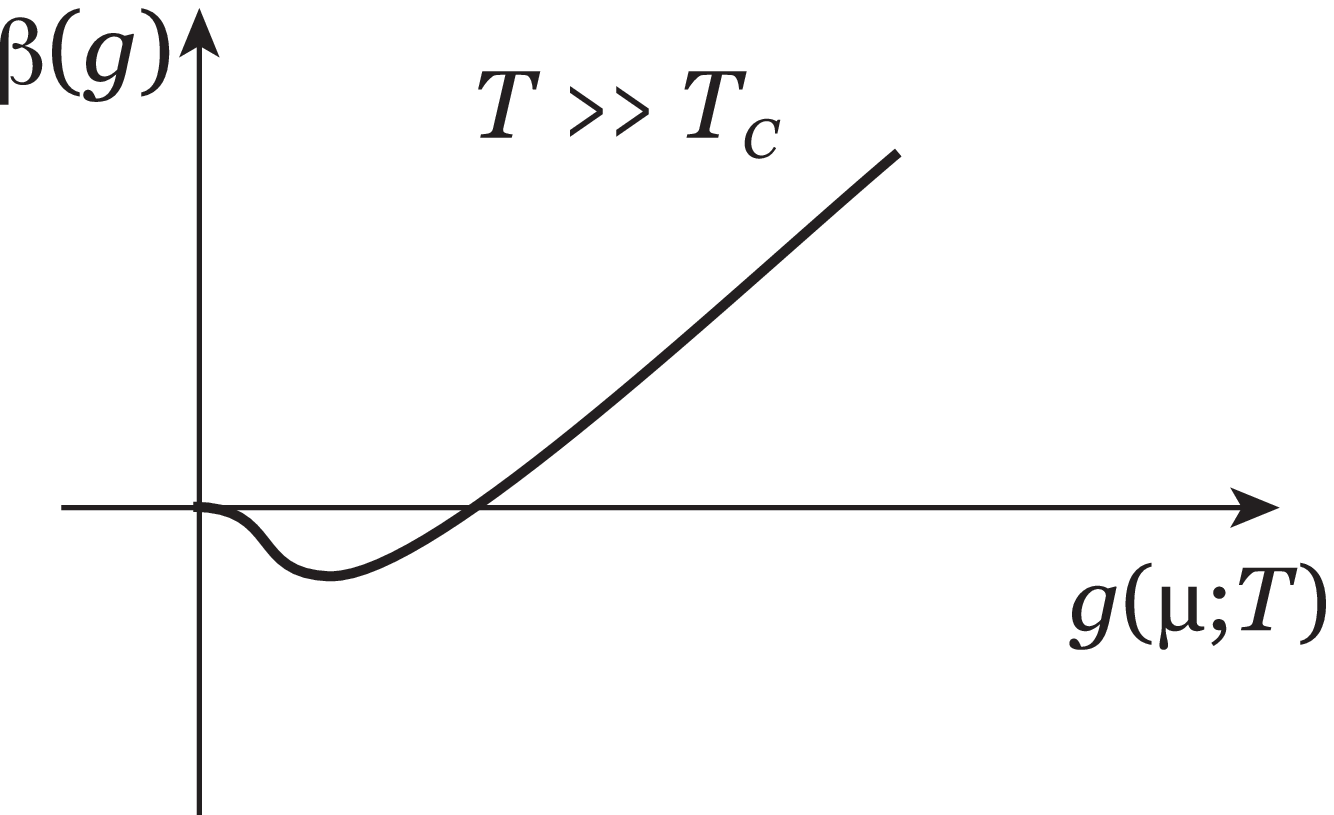,width=2in}\figsubcap{b}}
  \caption{$\beta$ function with $T=$ constant. (a) $T \ge T_c$. (b) $T \gg T_c.$}
  \label{beta_gT}
\end{center}
\end{figure}

As far as  $T < T_c,$ the $\beta$ function is negative all through $g$.
As the temperature is increased further, the form of $\beta$ function will changes
as in Fig.\ref{beta_gT}: (a) When $T > T_c$ but $T \sim T_c$, the $\beta$ function changes the sign from negative to positive at large $g$. (b) When $T \gg T_c$ it changes the sign at small $g$.

The $T/T_c$ dependence of the $\beta$ functions resembles the $N_f$ dependence~\cite{Banks1982}  for large 
$N_f$: $N_f=16$ $\Leftrightarrow$ $T/T_c \gg 1$ and $N_f=7$ $\Leftrightarrow$ $T/T_c \ge 1.$

\subsection{Numerical Results at HIgh Temperatures}
\label{numerical}

We have made simulation in the $N_f=2$ case on a $16^3\times 64$ lattice for several sets of parameters.
Let us discuss the three cases listed in Table.1:
$\beta=6.5$  corresponds to $T \sim 2\, T_c$,
$\beta=10.0$ to $T \sim 100 \, T_c$,
and $\beta=15.0$  to $T \sim 10^5\, T_c$.

We clearly see that the propagators in the three cases decay with the modified Yukawa type decay form.
Therefore we make the local-analysis of the propagators.
The $m(t)$ and $\alpha(t)$ are depicted in Figs.~\ref{Nf2K=146L16}, \ref{Nf2K=135L16} and  \ref{Nf2K=130L16}.

\section{Correspondence between $N_f$ and $T$}
\vspace{-0.2cm}
We observe the $t$ dependence of $m(t)$ and $\alpha(t)$ in the case $T\sim 2 T_c$
are very similar to those in the $N_f=7$ case. 
On the other hand, the $t$ dependence in the $T \sim 100 T_c$ and $T\sim 10^5 T_c$ cases 
are very similar to those in the $N_f=16$ case. 
We plot in Figs.~\ref{Nf7Nf2},  \ref{Nf16Nf2_1} and \ref{Nf16Nf2_2}, the large $N_f$ data on the right panel
and the $N_f=2$ at high temperature data on the left panel in order to compare them directly.

The similarities between them are excellent. 
We could anticipate this from the similarities of the $\beta$ functions:
 $N_f=16$ $\Leftrightarrow$ $T/T_c \gg 1$ and $N_f=7$ $\Leftrightarrow$ $T/T_c \ge 1.$
 
We point out a possibility that there is a correspondence between the large $N_f$ with $\Lambda_{IR}$ and small $N_f$ at high temperature $T$.
Both of the large $N_f$ with $\Lambda_{IR}$ and the small $N_f$ at high temperature $T$ are described by two
parameters in addition to $N$ and $N_t$; $\beta$ and $m_q.$
In order to have a correspondence we have to map from one set of the parameters to the other set.
Although It is too premature to discuss the details,
the correspondence between $N_f$ and $T$ would be very intriguing and very useful. Although the flavor number $N_f$ is a discrete integer, the temperature $T$ is a continuous number. 
Thus the properties of large $N_f$ could be deduced from those of small $N_f$ at high temperatures with this correspondence.

\section{Models}
We investigate a few models to compare the results with the models.
\vspace{-0.2cm}
\begin{enumerate}
\item
Meson unparticle model
\item 
Fermion unparticle model
\end{enumerate}
The meson unparticle model and fermion unparticle model are motivated by the soft-wall model in AdS/CFT correspondence \cite{Cacciapaglia:2008ns}.

\begin{enumerate}
\item
 The soft-wall model predicts the form of the propagator with scale dimension $\Delta$ in the momentum space as
\begin{equation}
\langle O(p) O(-p) \rangle = \frac{1}{(p^2+m^2)^{2-\Delta}} \ .
\end{equation}

In this case, when $m\, t \ll 1,$
$\alpha(t)=3-2 \, \gamma^{*}$. This is model independent and universal.
On the other hand, when $m\, t \gg 1,$
$\alpha(t)=2-\gamma^{*}$ for $t \gg \Lambda_{CFT}^{-1}.$ 
This is model dependent.

\item

Another possible scenario is to treat $\bar{\psi}\gamma_5 \psi(x)$ as the non-bound state of unfermions. 
The soft-wall model predicts the form of the propagator with scale dimension $\Delta_f$ in the momentum space as
\begin{equation}
\langle \Psi(p) \bar{\Psi}(-p) \rangle = (p^\mu \gamma_\mu + m) \frac{1}{(p^2+m^2)^{\frac{5}{2}-\Delta_f}} \ .
\end{equation}

In this case, when $m\, t \ll 1,$
$\alpha(t)=3-2 \, \gamma^{*}$.
On the other hand, when $m\, t \gg 1,$
$\alpha(t)=1.5-\gamma^{*}$ for $t \gg \Lambda_{CFT}^{-1}.$ 
This is model dependent.
\end{enumerate}

\section{What Kind of Theories are defined ?}
First let us consider the $N_f=16$ case. 
We expect that the $N_f=16$ case is close to a free fermion state.
After the conference we further investigate this issue, and find that it is indeed close to a free fermion state, but the vacuum
is the $Z(3)$ twisted vacuum.
 We also find that 
$N_f=2$ case at $T\sim 100\, T_c$ and $T\sim 10^5\, T_c$ is close to a free fermion state in the $Z(3)$ twisted vacuum.
The approach to a free state is very slow.
Investigation in detail of propagators reveals that the difference cannot be attributed to the fermion unparticle model.
This is different what we thought at the conference.

Next let us consider the $N_f=7$ case. The $t$ dependence of $\alpha(t)$ is very similar to the $N_f=2$ case at $T\sim 2\, T_c.$
The $\alpha(t)$ shows a plateau at large $t$ ($t=15\sim 31$).
We conclude from this fact that the $N_f=7$ theory is close to the meson unparticle model.

Now we estimate $\gamma^*$ for $N_f=7$, using the formula we have derived.
The value of $\alpha(t)$ at plateau($t=15\sim 31$) is $0.8(1)$ for $K=0.1452$ 
and $K=0.1459$.
Applying the formula $\alpha(t)=2 -\gamma^*$, we have $2 -\gamma^* = 0.8(1)$.
Thus $\gamma^* = 1.2(1).$ 
Although the estimate should be much more refined, taking the continuum limit, 
we may expect that  $\gamma^* $ is order of unity.

\section{Solutions for Long Standing Issues at High Temperatures}

There are long standing issues at high temperatures such as:

The free energy of quark-gluon plasma state does not reach that of the Stefan-Boltzmann ideal gas limit even at $T/T_c=100.$ \cite{fodor}

Wave functions of ¡Èmeson¡É just above $T_c$ can be obtained, although quarks are not confined.

What is the order of the chiral phase transition in the $N_f=2$ case ?  
This is theoretically and phenomenologically important issue.

We concentrate the issue of the order of the chiral phase transition in the $N_f=2$ case in this report.

Pisarski and Wilczek \cite{pisarski}
mapped $N_f=2$ QCD at high temperature 
to the three dimensional sigma model and 
pointed out that if $U_A(1)$ symmetry is not recovered at the chiral transition temperature,
the chiral phase transition of QCD in the $N_f=2$ case is 2nd order 
with exponents of the  three dimensional $O(4)$ sigma model.

The $O(4)$ scaling relation was first tested for staggered quarks~\cite{karsch94}.
For the Wilson quarks it was shown
that
the chiral condensate satisfies remarkably  the $O(4)$  scaling relation,
with the RG improved gauge action and the Wilson quark action \cite{iwa1997}  and 
with the same gauge action and the clover-improved Wilson quark action \cite{cppacs2001} (See, for example, Fig.6 in ref.~\cite{iwa1997}).
It was also shown for staggered quarks the scaling relation is satisfied 
 in the $N_f=2 + 1$ case \cite{staggered}, extending the region from $T/T_c >1$ adopted in
 \cite{iwa1997} and  \cite{cppacs2001}
 to the region including $T/T_c<1$.
 These results imply the transition is second order.
 
 However, recently, it was shown that the expectation value of 
 the chiral susceptibility $\chi_{\pi} -\chi_{\sigma}$ is
 zero~\cite{aoki2012} in thermodynamic limit
 when the $SU(2)$ chiral symmetry is recovered under the assumptions we will discuss below. This is consistent with that the  $U_{A}(1)$ symmetry is recovered, which
implies the transition is 1st order according to \cite{pisarski}. 
Apparently the  two conclusions are in contradiction.

Here we revisit this issue with the new insight of conformal field theories with an IR cutoff.
It is assumed in ref.~\cite{aoki2012}  that
the vacuum expectation value of mass-independent observable is an analytic function of 
$m_q^2$, if the chiral symmetry is restored.
However, in the conformal region
the propagator of a meson behaves as eq.(3) and the relation between the $m_H$ and the $m_q$ is given by the hyper-scaling relation~\cite{miransky} 
$$m_H = c \, m_q^{1/(1+\gamma*)},$$
with $\gamma^*$ the anomalous mass dimension.
This anomalous scaling implies $m_H$  is not analytic in terms of $m_q^2$ and  the analyticity assumption does not hold.
It should be noted that the Ward -Takahashi Identities
in~\cite{aoki2012} are proved in the thermodynamic limit and therefore
the numerical verification of the hyperscaling in the limit will be decisive.
We stress however that the hyper-scaling is theoretically derived with
the condition of the existence of the IR fixed point and
multiplicative renormalization of $m_q$. We believe that this
violation of the analyticity assumption resolves the apparent
discrepancy as also mentioned in~\cite{aoki2012} as a viable
possibility.

For the other issues given in the beginning of this section,
we only mention that the slow approach to the free fermion state in the $Z(3)$ twisted vacuum is related to the slow  
approach of the free energy of quark-gluon plasma state to the Stefan-Boltzmann ideal gas limit,
and
when $T \sim T_c$, ''mesons'' are meson unparticles, similar to meson particles in some aspects.
We will investigate these points further.

\section{Discussion and Conclusions}
First of all, our conjectures for the
¡Èconformal theories with an IR cutoff¡É  have been verified in the cases of $N_f=7$ and $N_f=16.$
This fact is consistent with 
the assertion that the conformal window is $7 \le N_f \le 16$.
We have further verified the
¡Èconformal theories with IR cutoff¡É in the case $T/T_c >1$ for $N_f=2.$
We have stressed that
IR cutoff is inherent with simulations on a lattice and QCD at high temperatures.

The local-analysis of propagators is proposed and is shown to be very useful. 
Even in the confining phase the $t$ dependence of
$\alpha(t)$ contains information of RG flow from UV to IR. If it is affected by an IR fixed point, it will appear as remnant in the $\alpha(t).$ 
If  it is not affected, it is difficult to estimate $\gamma^*$ correctly from the scaling rule,~eq.(\ref{scaling}).

By the ''local-analysis'' of the propagator, we observe 
''$N_f=7$'' and ''$T\sim T_c$'' are similar to each other, and ''$N_f=16$'' and ''$T\gg T_c$'' are similar to each other. 
Using the correspondence between ''$N_f$'' and ''$T$'' that our data suggest, we conclude our data are consistent with the picture that
''$N_f=7$'' and ''$T\sim T_c$'' are close to the meson unparticle model,
and estimate $\gamma^* = 1.2(1)$.

Surely a lot of things theoretically and numerically should be done.
In particular, the continuum limit of propagators in the large $N_f$ case should be estimated in order to obtain with more confident what kind of theories are defined. To do so, we have to investigate finite size effects and the $\beta$ dependence of propagators.
In parallel with these efforts, physics implications of unparticles such as the effects in the early universe should be much more deepened.

\section*{Acknowledgments}
I would express gratitude to K.-I. Ishikawa, Yu Nakayama and T. Yoshie for a stimulating and
fruitful collaboration. I also would like to thank T. Yanagida for making a chance to start this collaboration.
I thank K. Kanaya for his help in preparing the manuscript.
I would like to thank the organizer of SCGT 2013 for giving the chance to talk on this subject.
The calculations were performed using HA-PACS computer at CCS, University of Tsukuba and SR16000
at KEK. I would like to thank members of CCS and KEK for their strong support for this work.


\begin{figure}[b]
\begin{center}
\psfig{file=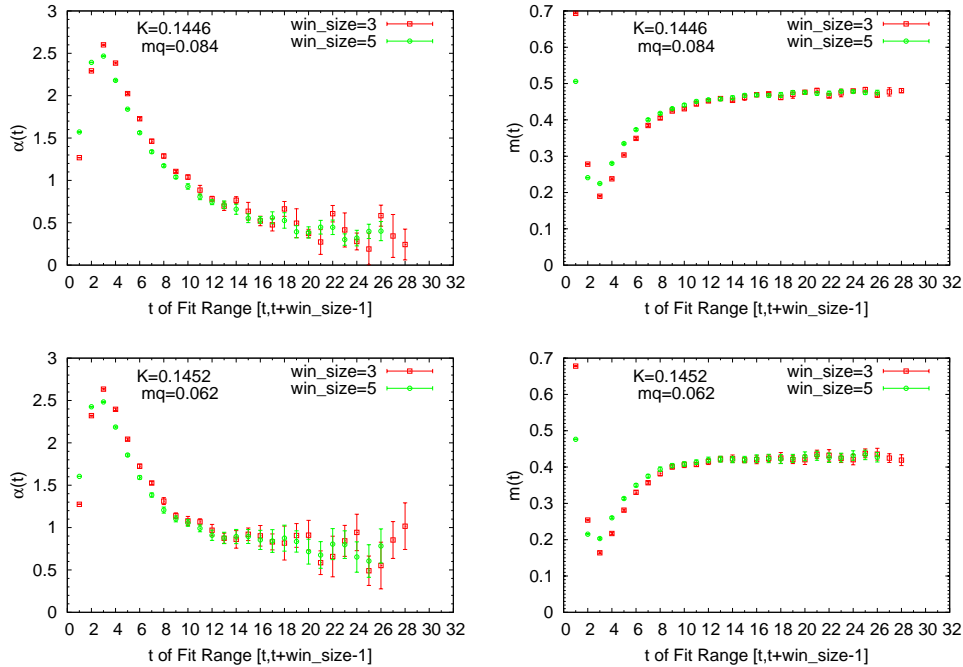,width=13.2cm}
\end{center}
\vspace{-0.7cm}
\caption{$\alpha(t)$ and $m(t)$; $N_f=7$ on a lattice $16^3\times 64$ at $K=0.1446$ and $K=0.1452$}
\label{K1446K1452L16}
\end{figure}

\begin{figure}[t]
\begin{center}
\psfig{file=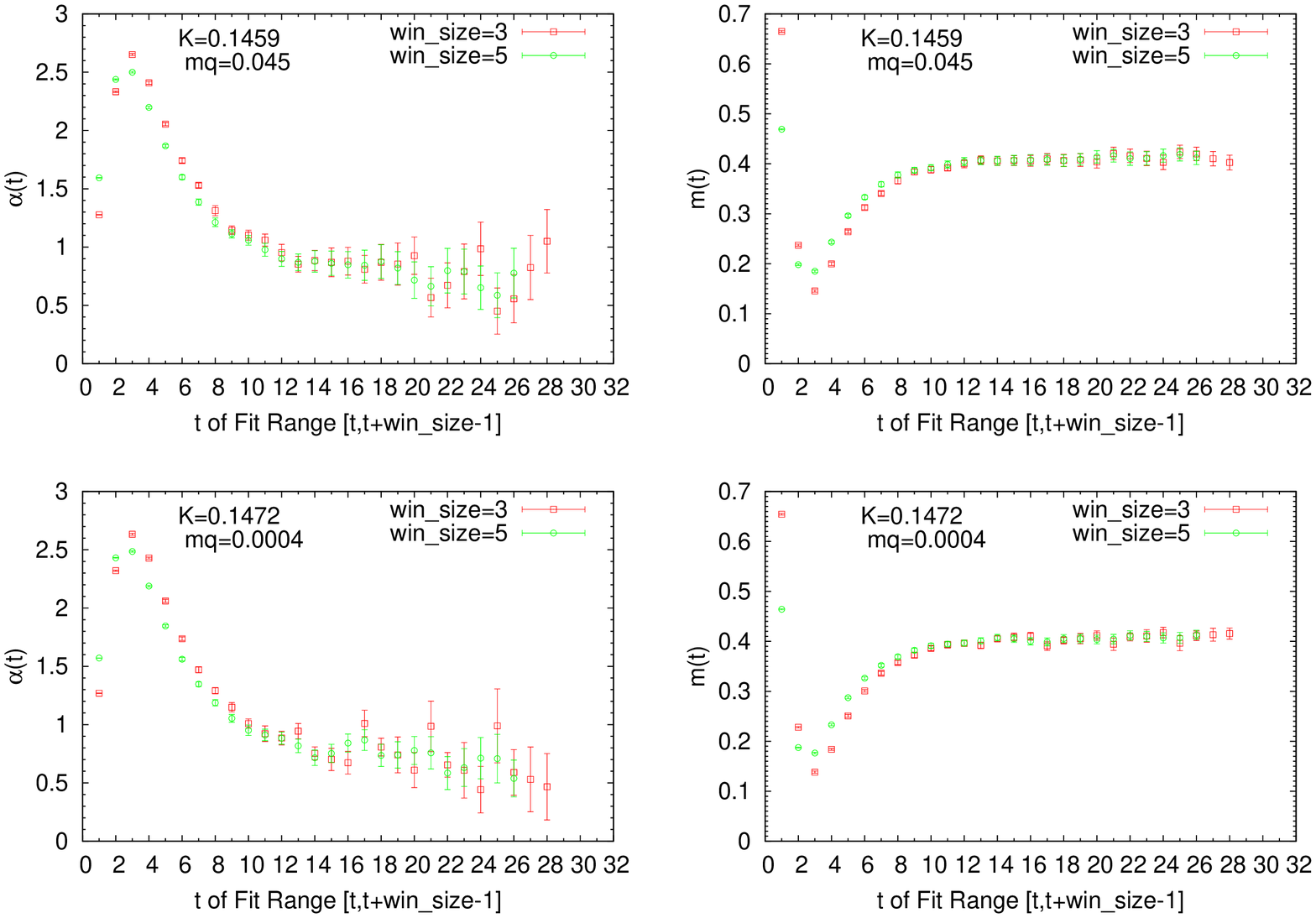,width=13.2cm}
\end{center}
\vspace{-0.7cm}
\caption{$\alpha(t)$ and $m(t)$; $N_f=7$ on a lattice $16^3\times 64$ at $K=0.1459$ and $K=0.1472$}
\label{K1446K1472L16}
\end{figure}

\vspace{-5.0cm}

\begin{figure}[b]
\begin{center}
\psfig{file=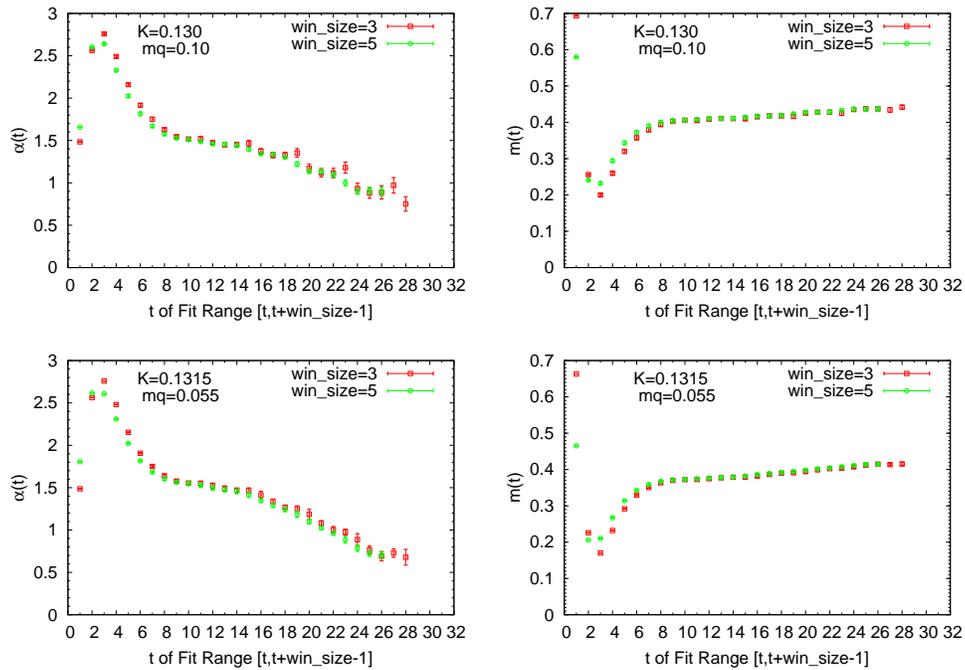,width=13.2cm}
\end{center}
\vspace{-0.7cm}
\caption{$\alpha(t)$ and $m(t)$; $N_f=16$ on a lattice $16^3\times 64$ at $K=0.130$ and $K=0.1315$}
\label{K130K1315L16}
\end{figure}

\vspace{-5.0cm}

\begin{figure}[t]
\begin{center}
\psfig{file=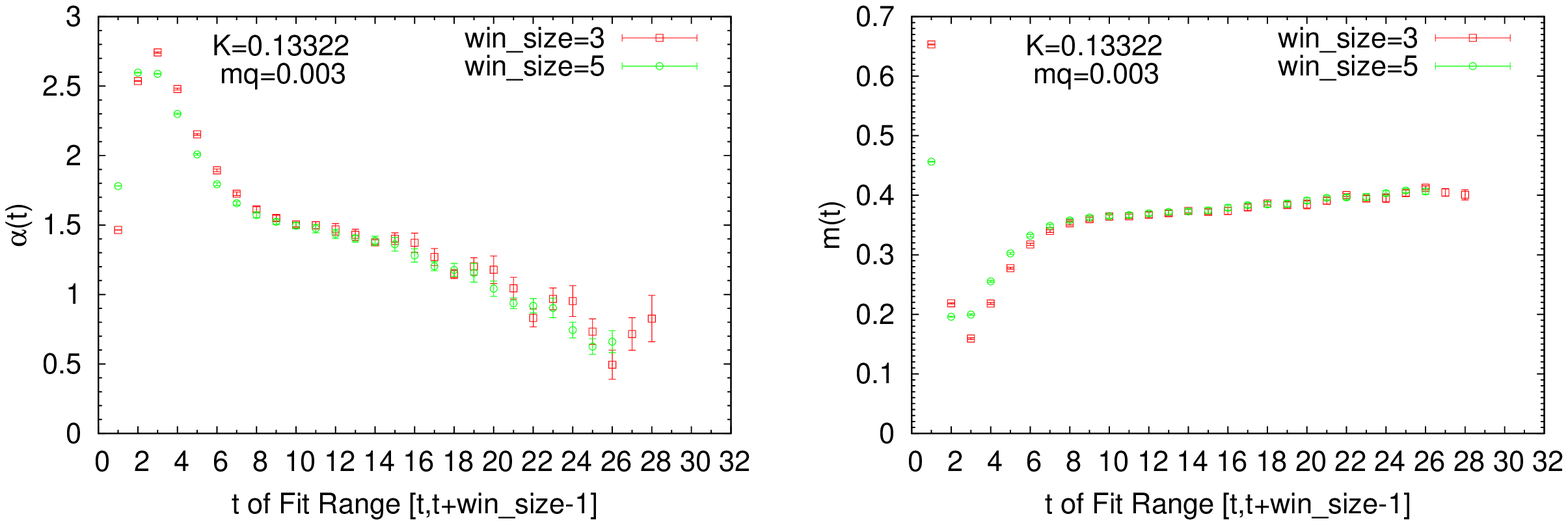,width=13.2cm}
\end{center}
\vspace{-0.7cm}
\caption{$\alpha(t)$ and $m(t)$; $N_f=16$ on a lattice $16^3\times 64$ at $K=0.13322$}
\label{K=0.13322L16}
\end{figure}

\vspace{-2.0cm}

\begin{figure}[t]
\begin{center}
\psfig{file=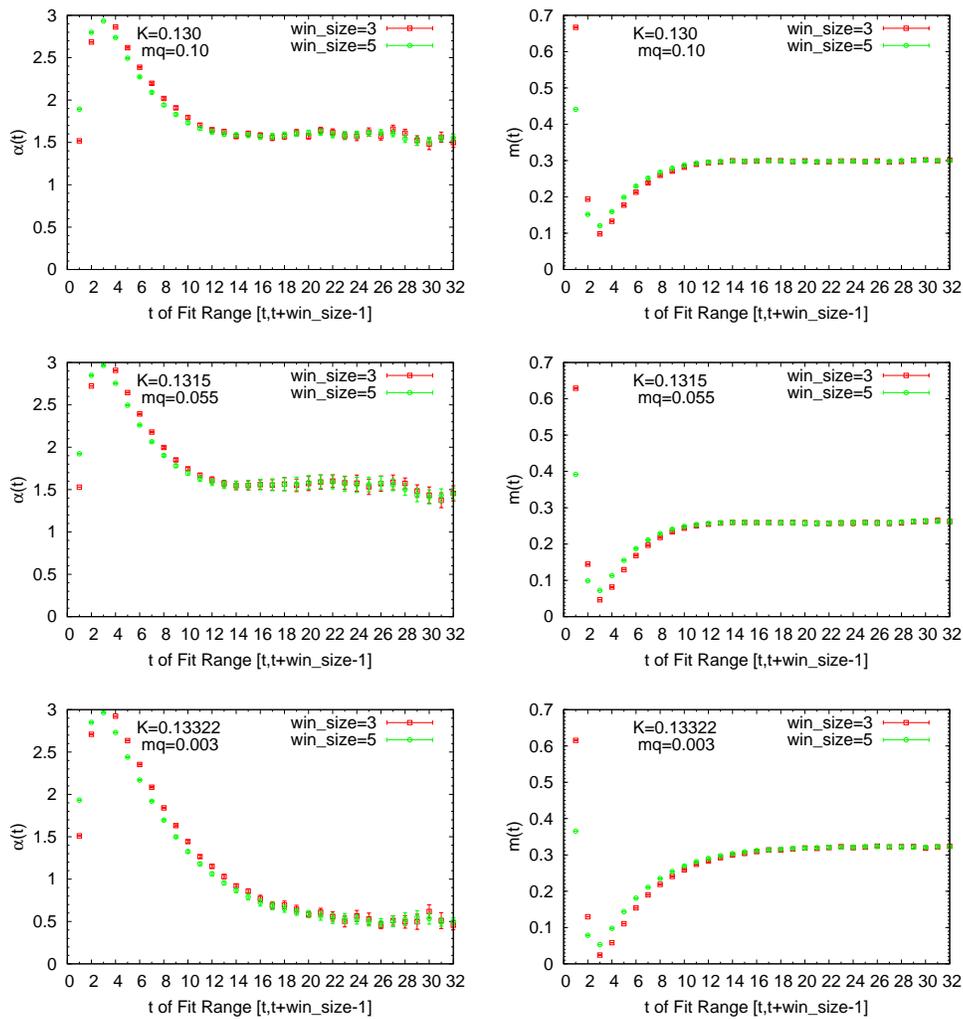,width=13.2cm}
\end{center}
\vspace{-0.7cm}
\caption{$\alpha(t)$ and $m(t)$; $N_f=16$ on a lattice $24^3\times 96$ at $K=0.130$, $K=0.1315$ and 
$K=0.13322$}
\label{K130131513322L24}
\end{figure}
\vspace{-2.0cm}

\begin{figure}[t]
\begin{center}
\psfig{file=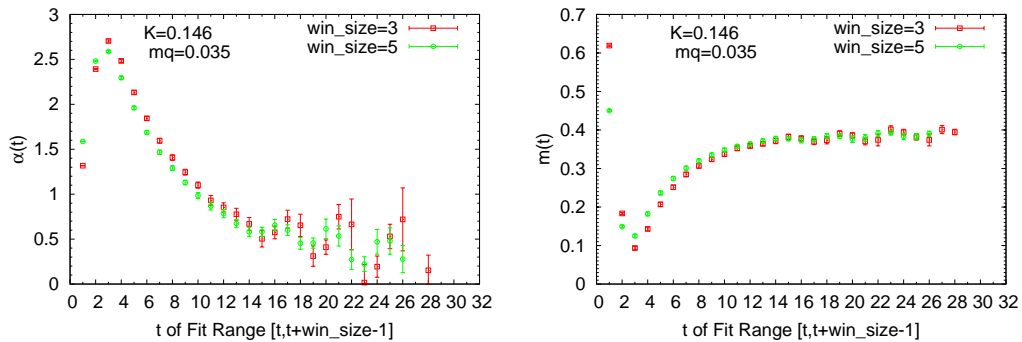,width=14cm}
\end{center}
\vspace{-0.7cm}
\caption{$\alpha(t)$ and $m(t)$; $N_f=2 $ on a lattice $16^3\times 64$ at $\beta=6.5$ and $K=0.146$.}
\label{Nf2K=146L16}
\end{figure}

\vspace{-2.0cm}

\begin{figure}[t]
\begin{center}
\psfig{file=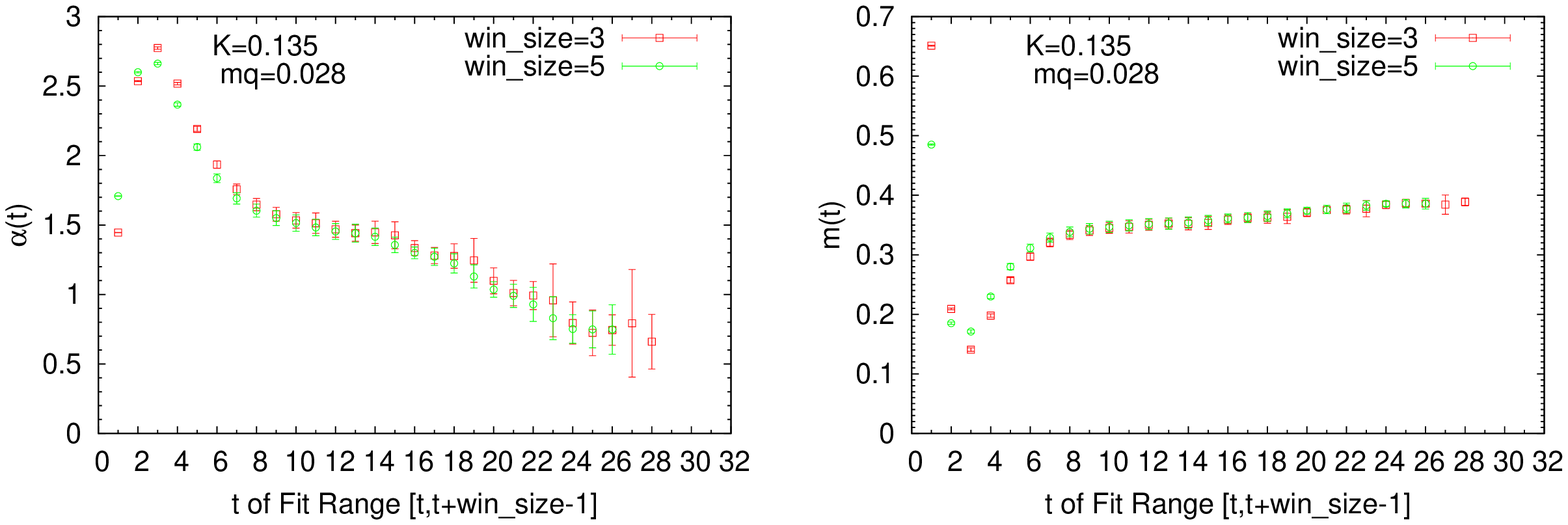,width=14cm}
\end{center}
\vspace{-0.7cm}
\caption{$\alpha(t)$ and $m(t)$; $N_f=2$ on a lattice $16^3\times 64$ at $\beta=10.0$ and $K=0.135$.}
\label{Nf2K=135L16}
\end{figure}

\vspace{-2.0cm}

\begin{figure}[t]
\begin{center}
\psfig{file=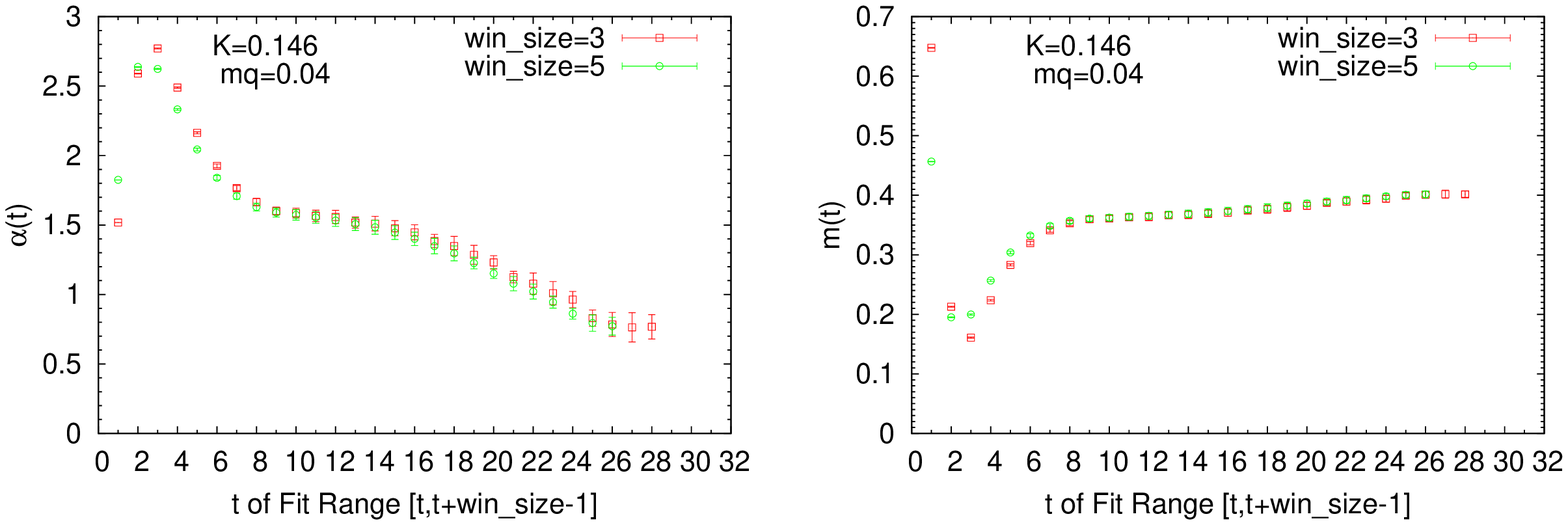,width=14cm}
\end{center}
\vspace{-0.7cm}
\caption{$\alpha(t)$ and $m(t)$; $N_f=2 $ on a lattice $16^3\times 64$ at $\beta=15.0$ and $K=0.130$.}
\label{Nf2K=130L16}
\end{figure}

\def\figsubcap#1{\par\noindent\centering\footnotesize(#1)}
\begin{figure}[b]%
\begin{center}
  \parbox{2.1in}{\epsfig{figure=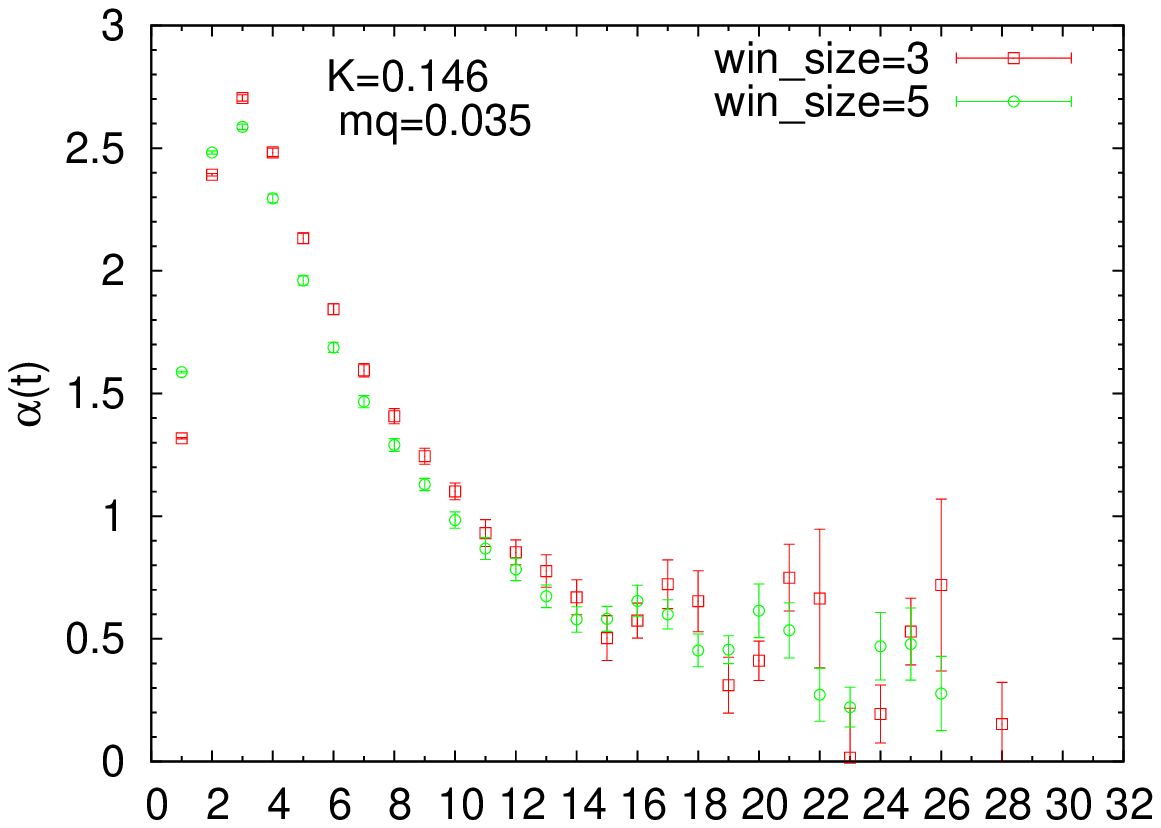,width=2.5in}\figsubcap{a}}
  \hspace*{20pt}
  \parbox{2.1in}{\epsfig{figure=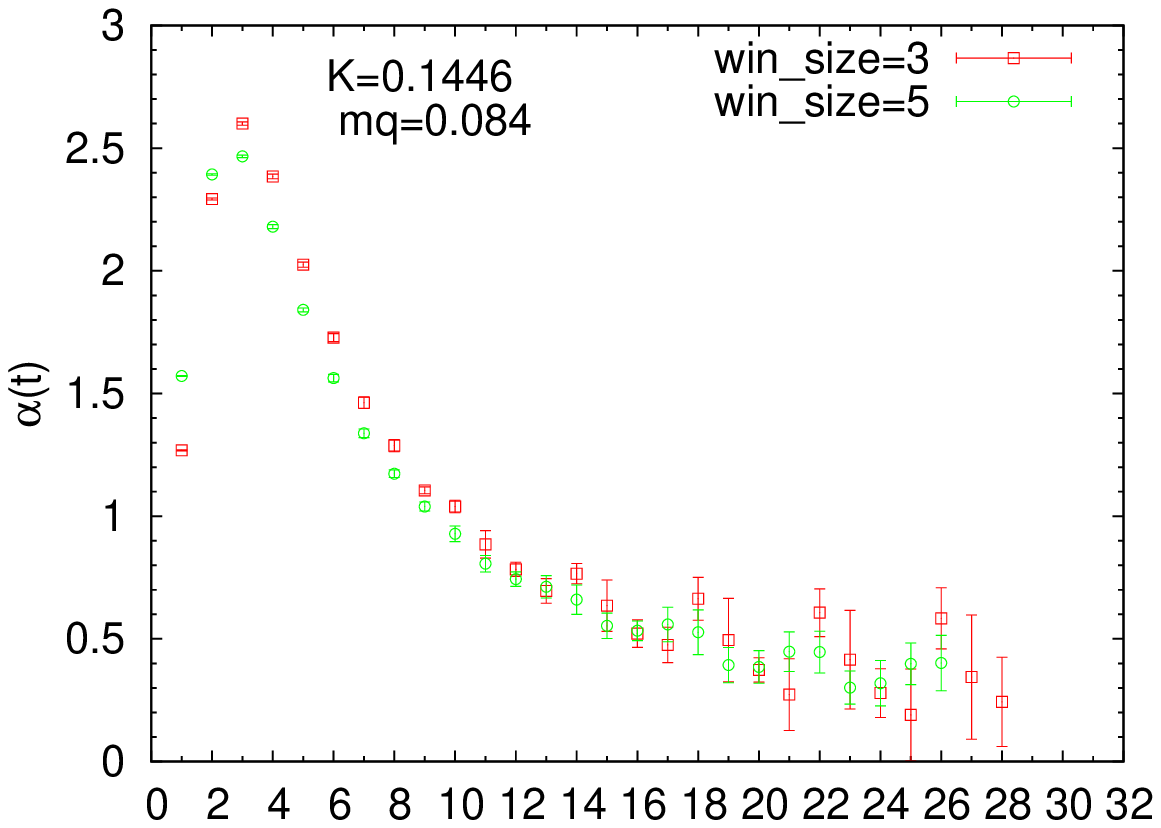,width=2.5in}\figsubcap{b}}
  \caption{The exponent $\alpha(t)$ on a lattice $16^3\times 64 $.(a) $N_f=2  T \sim 2\, T_c$.(b) $N_f=7; K=0.1446$.}
  \label{Nf7Nf2}
\end{center}
\end{figure}

\def\figsubcap#1{\par\noindent\centering\footnotesize(#1)}
\begin{figure}[b]%
\begin{center}
  \parbox{2.1in}{\epsfig{figure=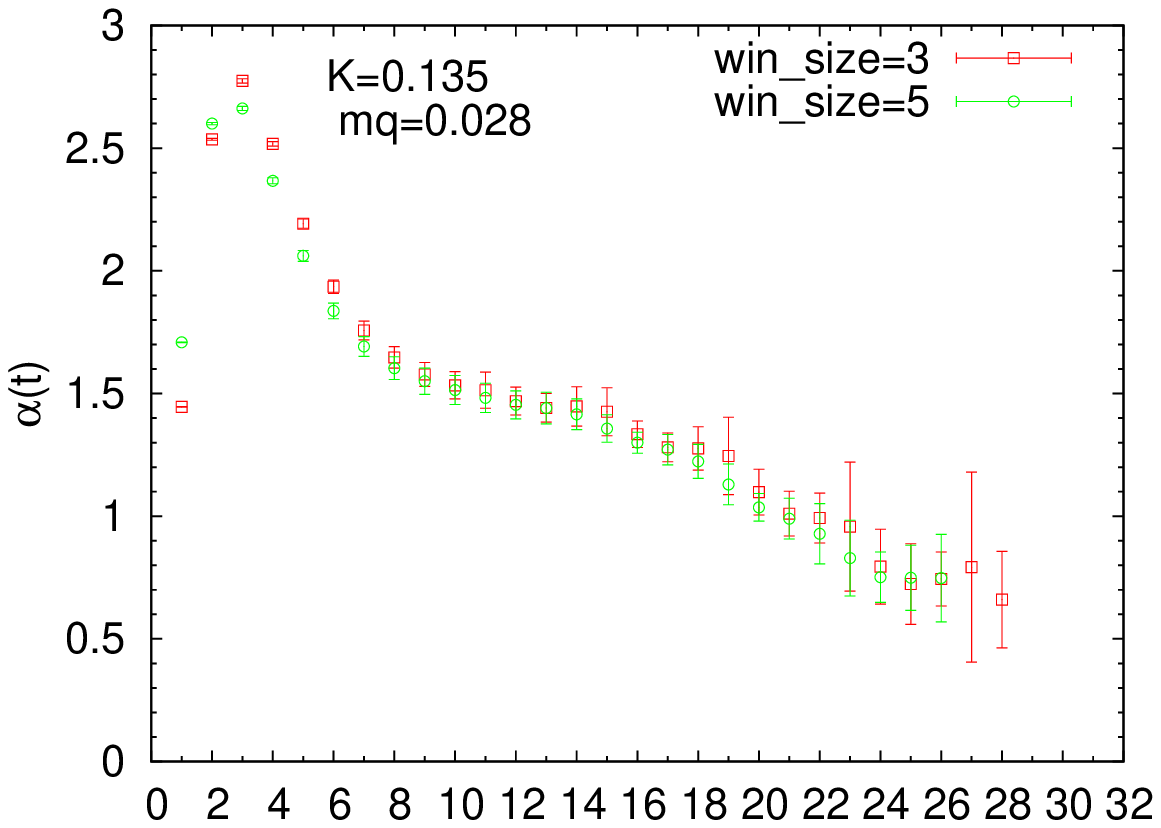,width=2.5in}\figsubcap{a}}
  \hspace*{20pt}
  \parbox{2.1in}{\epsfig{figure=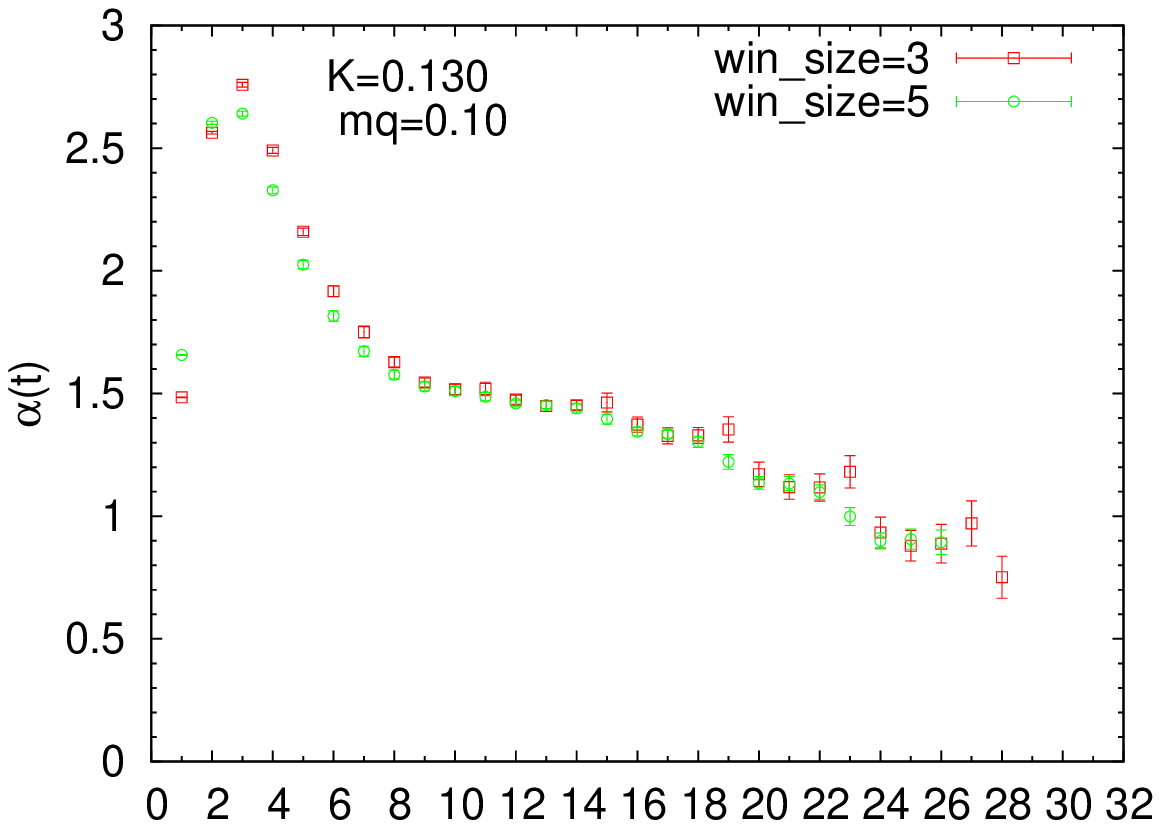,width=2.5in}\figsubcap{b}}
  \caption{The exponent $\alpha(t)$ on a lattice $16^3\times 64$. (a) $N_f=2;  T \sim 100\, T_c$. (b) $N_f=16; K=0.130$.}
  \label{Nf16Nf2_1}
\end{center}
\end{figure}

\def\figsubcap#1{\par\noindent\centering\footnotesize(#1)}
\begin{figure}[b]%
\begin{center}
  \parbox{2.1in}{\epsfig{figure=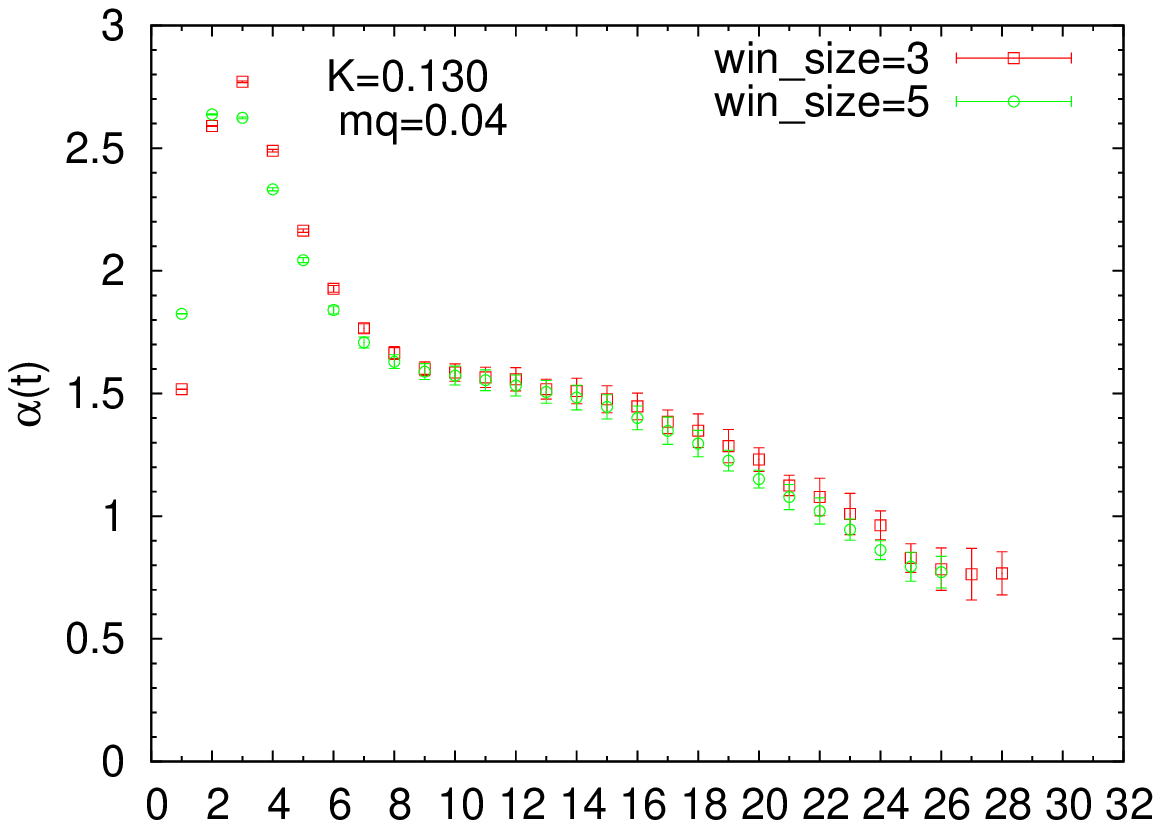,width=2.5in}\figsubcap{a}}
  \hspace*{20pt}
  \parbox{2.1in}{\epsfig{figure=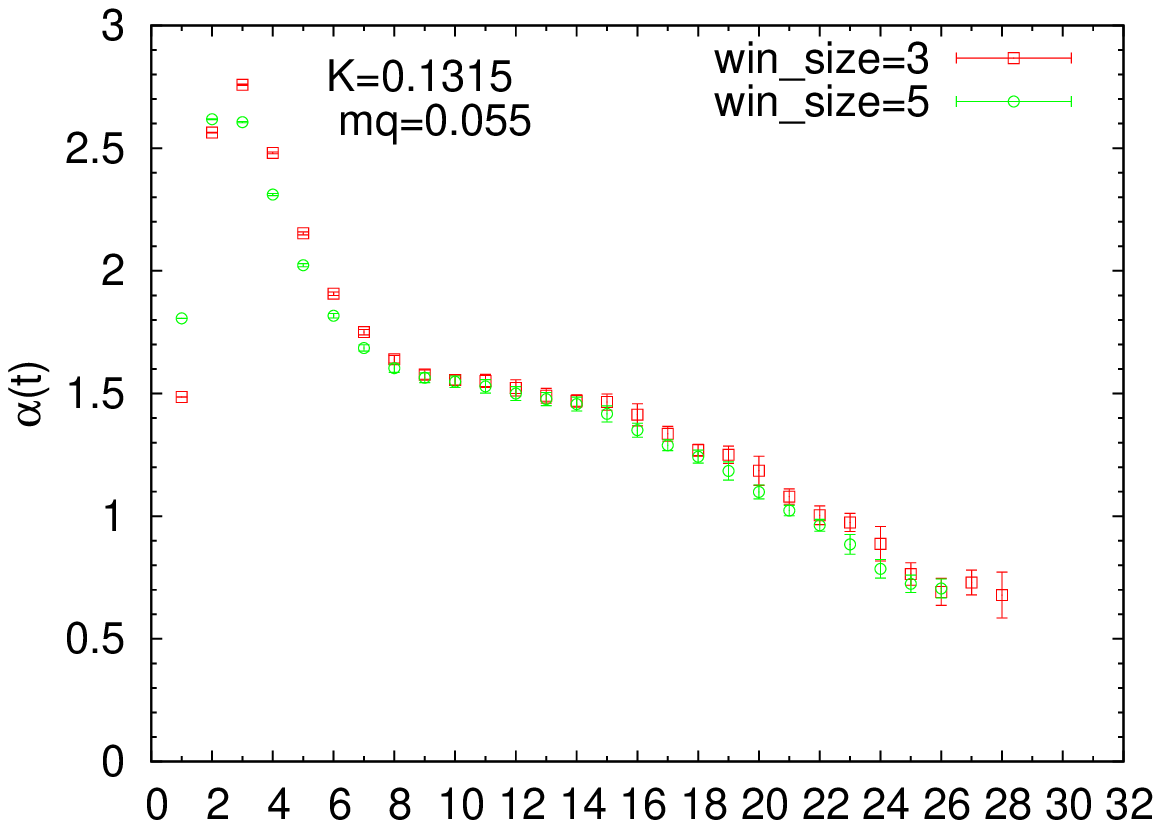,width=2.5in}\figsubcap{b}}
  \caption{The exponent $\alpha(t)$ on a lattice $16^3\times 64$. (a) $N_f=2; T \sim 10^5\, T_c$. (b) $N_f=16; K=0.1315$.}
    \label{Nf16Nf2_2}
\end{center}
\end{figure}


\end{document}